\documentclass[a4paper,fleqn,usenatbib]{mnras}

\usepackage[T1]{fontenc}
\usepackage{ae,aecompl}

\usepackage{graphicx}	% Including figure files
\usepackage{amsmath}	% Advanced maths commands
\usepackage{amssymb}	% Extra maths symbols

\title[A closer look at the ``characteristic'' width of filaments]{A closer look at the ``characteristic'' width of molecular cloud filaments}

\author[G. V. Panopoulou et al.]{
G. V. Panopoulou,$^{1,2}$\thanks{E-mail: panopg@physics.uoc.gr}
I. Psaradaki,$^{1}$
R. Skalidis, $^{1}$
K. Tassis,$^{1,2}$
J. J. Andrews$^{2}$
\\
$^{1}$Department of Physics and ITCP\thanks{Institute for
  Theoretical and Computational Physics, formerly Institute for Plasma
Physics}, University of Crete, 71003, Heraklion, Greece\\
$^{2}$Foundation for Research and Technology - Hellas, IESL, Voutes, 71110 Heraklion, Greece\\
}

\date{Accepted 2016 November 22. Received 2016 November 22; in original form 2016 September 13}

\pubyear{2016}

\begin{document}
\label{firstpage}
\pagerange{\pageref{firstpage}--\pageref{lastpage}}
\maketitle

\begin{abstract}
Filaments in \textit{Herschel} molecular cloud images are found to exhibit a ``characteristic width''.
This finding is in tension with spatial power spectra of the data, which show no indication of this characteristic scale.
We demonstrate that this discrepancy is a result of the methodology adopted for measuring filament widths.
First, we perform the previously used analysis technique on artificial scale-free data, and obtain
a peaked width distribution of filament-like structures. Next, we repeat the analysis on three \textit{Herschel} maps and 
reproduce the narrow distribution of widths found in previous studies $-$ when considering the average width of each filament. 
However, the distribution of widths measured at all points along a filament spine is broader than the distribution of mean 
filament widths, indicating that the narrow spread (interpreted as a ``characteristic'' width) results from averaging. 
Furthermore, the width is found to vary significantly from one end of a filament to the other.
Therefore, the previously identified peak at 0.1 pc cannot be understood as representing the typical width of filaments. 
We find an alternative explanation by modelling the observed width distribution as 
a truncated power-law distribution, sampled with uncertainties. The position of the peak is connected to the lower truncation 
scale and is likely set by the choice of parameters used in measuring filament widths. We conclude that a ``characteristic'' 
width of filaments is not supported by the available data.
\end{abstract}


%%%%%%%%%%%%%%%%%%%%%%%%%%%%%%%%%%%%%%%%%%%%%%%%%%

%%%%%%%%%%%%%%%%% BODY OF PAPER %%%%%%%%%%%%%%%%%%

\section{Introduction}

Studies of the structure of molecular clouds can provide clues on how gas accumulates to form stars.
Gas in molecular clouds is found to be ordered in filamentary structures, a result highlighted especially by \textit{Herschel} 
observations of dust emission in nearby clouds \citep{andre2010}. Dense, self-gravitating filaments are often found to be 
co-spatial with young stars and dense prestellar cores \citep[e.g.][]{hartmann2002, andre2010, polychroni2013, konyves2015}, and 
hence may be important for understanding star formation. 

One of the most striking results from analyses of \textit{Herschel} data is that filaments in the Gould Belt clouds are found to exhibit 
a narrow distribution of average cross-sectional widths\footnote{Throughout this paper the term 'width' refers to the FWHM of a 
Gaussian fit to the inner-most part of a filament radial profile, the same definition used by \cite{arzoumanian2011}.}
\citep{arzoumanian2011}. This sharply peaked distribution 
\citep[with a mean at $\sim 0.1$ pc and with $\sim$ 70\% of values within 0.06$-$0.14 pc,][]{arzoumanian2011, koch2015} 
contains filaments spanning more than two orders of magnitude in column density.

This finding seems to contradict the expectation that filaments should contract (due to gravity) and hence increase in 
density while decreasing in radius. Though the existence of this characteristic scale is still poorly understood, it has 
been suggested that it must be connected to some physical mechanism, perhaps one involved in filament formation 
\citep{arzoumanian2011,andre2014}.
Qualitative arguments have connected this characteristic scale to the transition from supersonic to 
trans-sonic turbulence \citep{arzoumanian2011} and to the ambipolar diffusion length scale 
\citep[for both gravitationally unbound and bound structures][]{hennebelle2013, hennebelleandre}. 
Simple analytical models propose that the independence of filament width from column density may be a result of the balance
between accretion onto the (self-gravitating) filaments and dissipation of the turbulence within them \citep{hennebelleandre, heitsch2013}.
\cite{fischera} offered pressure confinement of isothermal cylinders as a possible model for 
self-gravitating filaments.
Most recently, \cite{basu} proposed a model in which filaments are magnetic ribbons, produced by large-scale, 
trans-Alfv\'{e}nic turbulent flows in a strong magnetic field. Their model is able to reproduce average widths that vary within a factor of two across two orders of magnitude in column density.
However, simulations which either include self-gravity and neglect magnetic fields or vice versa have yet to reproduce the observed 
distribution and independence on column density \citep{smith2014, ntormousi2016}. \cite{federrath2016} simulated isothermal,
self-gravitating, magnetized clouds with super-Alfv\'enic driven turbulence. His finding is that filament widths are peaked at 0.1 pc 
and appear constant for one order of magnitude in column density, when turbulence is operating. 
The proposed explanation is that the characteristic width is set by the
dissipation of turbulence in shocks. His model, however, fails to reproduce the correlation between filament and 
magnetic field orientations found in molecular clouds with \textit{Planck} \citep{planckXXXV}. 

One particularly puzzling observation regarding the apparent characteristic width of filaments
is the absence of its imprint on the spatial power spectra 
of \textit{Herschel} cloud images \citep{miville2016}. The spatial power spectrum of the 250 $\mu$m map of 
the translucent non-star-forming Polaris Flare is well fit by a power law from $\sim 2$ pc to $\sim 0.02$ pc \citep{miville}. 
At the same time, the distribution of filament widths in this cloud is found to have a prominent peak at 0.05 $-$ 0.07 pc
\citep[][]{arzoumanian2011,panopoulou2016}.
A similar situation is found in the Chamaeleon molecular cloud complex, where filament widths are peaked around 0.12 pc with a spread
of 0.04 pc \citep{deoliveira}. However, these authors find no indication of a typical filament width in the $\Delta$-variance 
spectra \citep{stutzki1998} of the clouds, even though the signature of cores and clumps is easily identified as a change in 
the slope of the $\rm \Delta$-variance spectrum at the corresponding size scales.

Motivated by this apparent discrepancy, in this work we retrace the steps in the analyses of filament width distributions.
In section \ref{sec:methods}, we briefly describe the analysis used for constructing the distribution of filament widths.
We first perform this analysis on an artificial filamentary image with no characteristic scale and find the distribution of
widths to have a broad peak (section \ref{sec:fractal}). We then repeat the analysis on \textit{Herschel} data of three clouds 
(the Polaris Flare, Aquila and IC 5146, section \ref{sec:distros}), showing that the narrow spread of the distribution of widths found in previous studies is likely a consequence of averaging along filaments. 
The constancy of filament widths may therefore not be inferred 
from this spread. Furthermore, we investigate the position of the peak of the distribution of widths 
and find that it is likely a result of the choice of range within which the filament width has been measured 
(section \ref{sec:explanation}).
Finally, we summarize our results in section \ref{sec:summary}.

\section{METHODS}
\label{sec:methods}

In order to reproduce the distributions of filament widths for the three clouds presented in \cite{arzoumanian2011} (the Polaris Flare,
Aquila, and IC 5146), we follow an analysis similar to their study. We use primarily the \textit{Herschel} Spectral and Photometric Imaging 
Receiver (SPIRE) maps of these clouds at 250 $\mu$m, unless explicitly stated otherwise in the text.

First, we employ the Discrete Persistent Structures Extractor \citep[{\tt DISPERSE}, ][]{sousbie2011}, 
to identify the filamentary structures in each image. 
{\tt DISPERSE} analyses the topology of a given map and extracts its skeleton, which corresponds to the ridges of elongated structures.
We select the parameters of {\tt DISPERSE} so that the resulting skeletons are visually similar to those shown in the previous studies of 
the three clouds \citep[][IC 5146, figure 3]{arzoumanian2011}, \citep[][Aquila, figure 3]{konyves2015}, \citep[][Polaris Flare, figure 1]
{andre2014}. The values of the parameters used are given in appendix \ref{sec:skeletons}.

Next, we provide the skeleton of {\tt DISPERSE} and the corresponding \textit{Herschel} image of each cloud as input to the 
Filament Trait-Evaluated Reconstruction ({\tt FILTER}) code\footnote{The code is available at: {\tt https://bitbucket.org/ginpan/filter}} 
\citep{panopoulou2014}. 
The objective of {\tt FILTER} is twofold: First, it post-processes the skeleton of {\tt DISPERSE} to only include continuous, 
non-spurious, structures (e.g. peaked well above the noise level). 
This is done by taking cross-sections at every pixel along the filament ridge and assessing each intensity profile.
Profiles that are not peaked around the filament ridge and above the noise-level are rejected. Second, {\tt FILTER} measures the width 
of each intensity profile along a filament. The width is defined as the FWHM of a Gaussian fit (with offset) to the inner-most part of the 
profile \citep[as in][]{arzoumanian2011}. In order to find this value automatically for every profile, Gaussians are fit iteratively 
to smaller and smaller distances from the filament ridge. The initial range used for fitting is a free parameter of the algorithm.
As has been shown by \cite{smith2014}, the distance up to which a Gaussian is fit is crucial in the determination of the width, 
as at larger distances the fit tends to miss the inner-most part of the profile. In each section we shall state explicitly which
starting value for the fitting range has been used. 
At the end of the iteration, the most frequent FWHM is assigned as the width of the profile, after deconvolution from 
the beam size\footnote{The most frequent FWHM has been found through tests (on data and artificial images) during code development to be a robust estimator of the width.}. 
The deconvolved width is found \citep[as in][]{konyves2015} by $\rm FWHM_d = \sqrt{FWHM^2 - HPBW^2}$, where HPBW is the half-power beam 
width of the map (18\arcsec for 250 $\mu$m, which corresponds to 0.013 pc, 0.023 pc, and 0.04 pc at the assumed distances to the Polaris Flare, 150 pc, Aquila, 260 pc, and IC 5146, 460 pc, respectively). Finally, only sufficiently elongated structures (with at least a 3:1 length to mean width ratio) are included in 
the final sample of filaments that we consider for further analysis. The post-processed skeletons of the three 
\textit{Herschel} images used in this work are shown in appendix \ref{sec:skeletons}.

{\tt FILTER} provides two different ways of constructing width distributions:
\begin{itemize}
\item[i.] A distribution of the FWHM measured at all points along the spines of filaments can be created. 
Information on the structure to which each measurement belongs is discarded in this way \citep[e.g.][]{panopoulou2016}.
We refer to this kind of distribution as the distribution of all profile widths.
\item[ii.] A distribution of the mean FWHM of filaments can be constructed. This second type of 
distribution has been found to show a ``characteristic width'' in previous studies \citep[e.g.][]{arzoumanian2011}.
We refer to this type of distribution as the distribution of filament-averaged widths (or of mean filament widths).
\end{itemize}  

With {\tt FILTER} the mean width of each filament is found by taking the average value of the FWHM measured at each point 
(pixel) along the filament ridge. 
In other studies, the mean width has been found by fitting a Gaussian to the mean filament profile
e.g. \cite{arzoumanian2011, koch2015, smith2014, benedettini}. In the work of \cite{koch2015}, a non-parametric 
width determination is also used when Gaussian fitting is not possible. In the following analysis we find the mean width of a filament by 
averaging the FWHM of all its profiles, as this process is automated, objective, and easily reproducible.
We note that the method used to calculate the average width of a filament should not affect the 
statistical properties of the distribution of mean filament widths (mean and spread), as long as a large number of filaments is used. 
Indeed, the difference between our approach of measuring mean filament widths and that of other studies does not affect
the properties of the distributions of mean filament widths, as both the mean and spread are in good agreement with 
those of \cite{arzoumanian2011} (section \ref{sec:distros}).

\begin{figure*}
\centering
\includegraphics[scale=1]{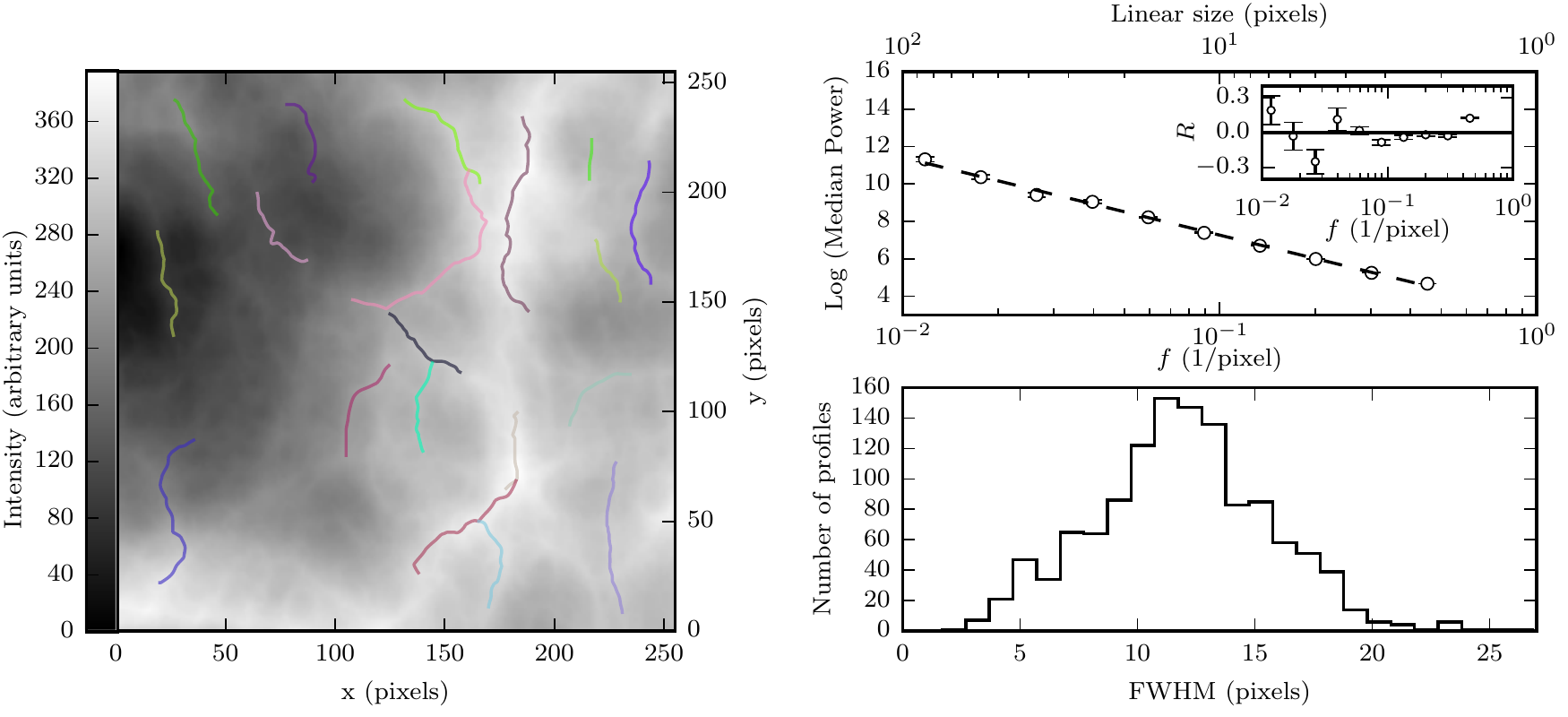}
\caption{Left: Image generated using Ridged Multifractal Noise (256 pixels on each side), with the skeleton of filaments 
having an aspect ratio of at least 3:1 overplotted. Top right: Spatial (azimuthally-averaged) power spectrum of image on left (open 
circles) and linear fit in log-log space (dashed line). Corner inset: Residual of the fit, $R =$ log(median power) - log(fit). Bottom right: Distribution of the widths measured at each point along the ridges of the 
filaments in the artificial image on the left.}
\label{fig:fractal_plots}
\end{figure*}

\section{RESULTS} 
\label{sec:results}

\subsection{Can a peaked size distribution arise from scale-free data?} 
\label{sec:fractal}

\begin{figure}
\centering
\includegraphics[scale=1]{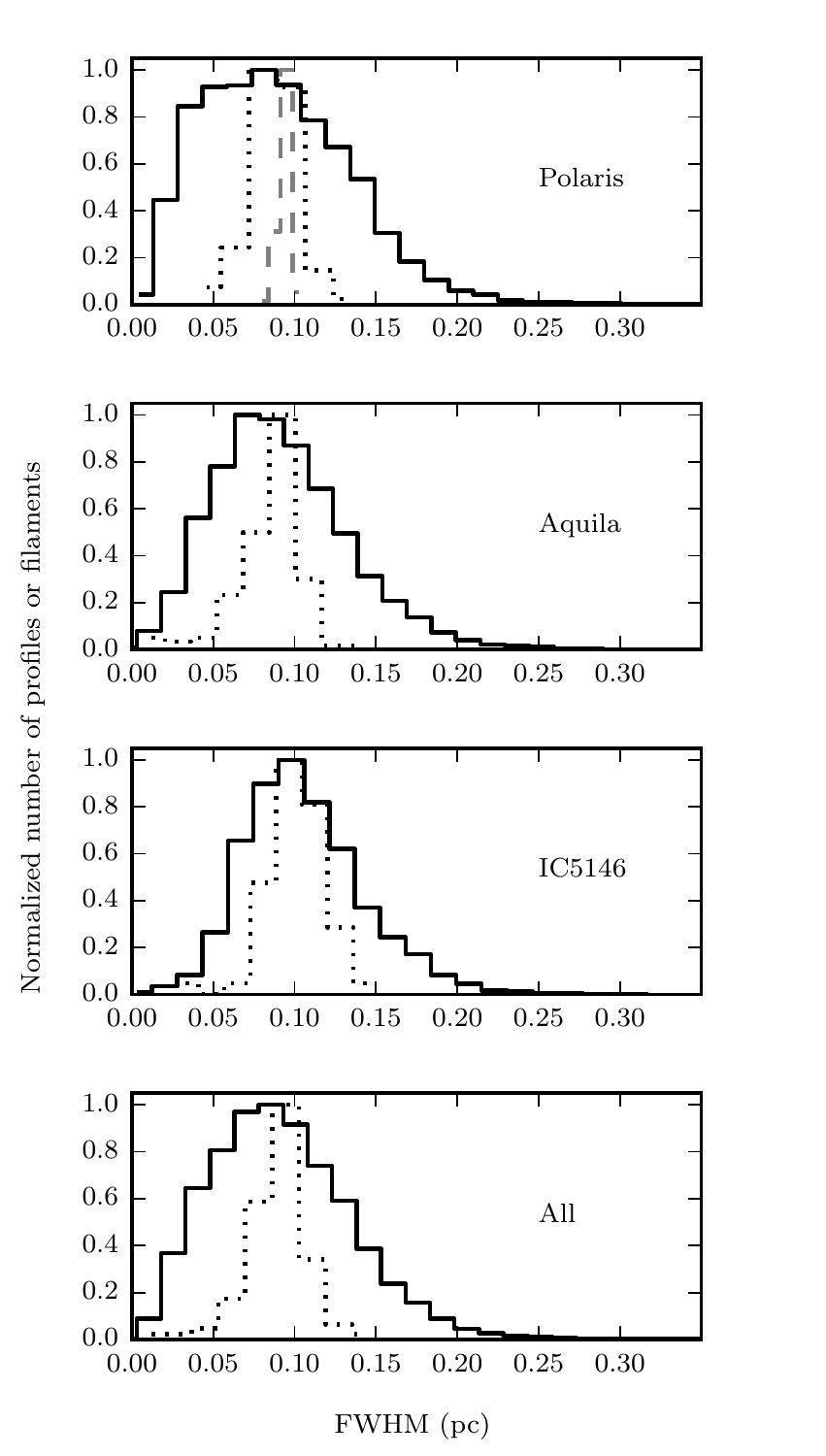}
\caption{Comparison of filament-averaged (dotted) and non-averaged (solid black) width distributions for the three different clouds (top 3 plots), and for all filaments in the three clouds combined (bottom). All FWHM have been deconvolved from the beam size. The grey dashed line (top panel) shows the distribution of mean filament widths resulting from the Monte Carlo simulation described in the text (section \ref{sec:distros}).}
\label{fig:all_distr}
\end{figure}

The scale-free spatial power spectrum of the \textit{Herschel} Polaris Flare image (at 250 $\mu$m)
is in tension with the existence of a ``characteristic'' width of the filaments in the same image \citep{miville,arzoumanian2011}.
It may be argued that in some circumstances, the imprint of a characteristic scale on the spatial power spectrum is ``hidden''. 
We use simple artificial images to explore if such a situation may arise, in appendix \ref{sec:powerspectra}. 
We demonstrate that if structures with a characteristic scale are introduced, an imprint of this scale is apparent
in the spatial power spectrum $-$ as long as the structures are easily discernible from background noise,  
as is the case for filaments in the Polaris Flare. 

In this section we explore the opposite situation: whether a preferred scale can arise from analysing an image with a scale-free spatial 
power spectrum.
To this end, we create a scale-free image and perform the analysis described in section \ref{sec:methods}.

To construct the image we use the Ridged Multifractal model from the suite of noise-synthesis models implemented in the python library
{\tt pynoise} ({\tt http://pynoise.readthedocs.io}). Noise synthesis models \citep{musgrave} are used for creating natural-looking complex 
and heterogenous patterns (landscapes, clouds). They are based on the widely used, scale-free, fractional-Brownian-motion 
(fBm) \citep[e.g.][]{stutzki1998} but use band-limited Perlin noise \citep{perlin} functions for the basis function 
instead of sine waves as in pure fBm. 

The resulting image (256 pixels on each side) is filamentary, as can be seen in Fig. \ref{fig:fractal_plots} (left). 
We adjusted the parameters of the model to obtain an image whose spatial (azimuthally-averaged) power spectrum 
has the form of a power law (Fig. \ref{fig:fractal_plots} - top right). 
The one dimensional spatial power spectrum is constructed as in \cite{pingel}, by taking the median power in concentric annuli around the 
zero-frequency pixel in the two-dimensional power spectrum. 
The annuli are chosen so that a uniform sampling of scales in logarithmic space is obtained. 
The sample size ranges from 29 (for the smallest annulus) to $\sim$23000 values.
The errors on the median value (comparable to the size of the points in the figure)
are calculated by bootstrap resampling. For every annulus, we resample the distribution of intensities and calculate the median of 
the resampled distribution 100 times. The error on the median is the standard deviation of these 100 median values.

We performed the analysis described in section \ref{sec:methods} on the artificial image. The resulting skeletons of filament-like 
structures (having at least a 3:1 aspect ratio) are overplotted in the left panel of Fig. \ref{fig:fractal_plots}. 
The distribution of widths measured at each point along these elongated structures (first method) is shown in the bottom right 
panel of Fig. \ref{fig:fractal_plots}. The initial fitting range used is $\pm$10 pixels from the ridge (see discussion in
section \ref{sec:methods} regarding the fitting range). 
The distribution is clearly peaked around 12 pixels, and has a spread of 3.8 pixels. 

The existence of a peak in the distribution of widths of the structures extracted by {\tt DISPERSE} in this image is 
inconsistent with its scale-free (power-law) spatial power spectrum.  
Since the construction of the power spectrum is straight-forward, we conclude that the existence of the preferred scale (peak of the
width distribution) is most likely an artefact of the analysis of apparent filament widths.

\subsection{Why is the distribution of widths narrow?} 
\label{sec:distros}

\begin{table}
\centering
\caption{Properties of width distributions shown in Fig. \ref{fig:all_distr}. 
Includes the number of filaments in each distribution, $N_{fil}$, the number of profiles of all filaments, $N_{pr}$,
the mean and standard deviation of the distribution of filament-averaged widths 
($\left\langle W_{mean} \right\rangle$ and $\sigma_{mean}$), 
and those of the distribution of all profile widths ($\left\langle W_{all} \right\rangle$, $\sigma_{all}$).}
\begin{tabular}{|c|c|c|c|c|c|c|}
\hline
Cloud & $N_{fil}$  & $N_{pr}$ & $\left\langle W_{mean} \right\rangle$ & $\left\langle W_{all} \right\rangle$& $\sigma_{mean}$& $\sigma_{all}$  \\
      &   &        &   (pc)                                & (pc)                                & (pc)            &  (pc) \\
\hline
\hline
Polaris & 100  & 24969 & 0.095 & 0.097 & 0.014 & 0.05 \\
\hline
Aquila  & 79  & 14315 & 0.095 & 0.094 & 0.02 & 0.04         \\
\hline
IC 5146 & 58  & 5277 & 0.11 & 0.11 &  0.02 & 0.04  \\
\hline
All		& 237 & 44561  & 0.09 & 0.09 & 0.02 & 0.04   \\
\hline
%\footnotesize{$^*$ $\rm s_{los} = N_H/n_H$.}
\end{tabular}
\label{tab:Nfils}
\end{table}

We now turn to the \textit{Herschel} data of the Polaris Flare, Aquila, and IC 5146. 
When combining width measurements of filaments in these clouds, \cite{arzoumanian2011} find a distribution of mean filament widths 
with a spread of only 0.03 pc. It is this small spread that seems to imply that filaments have a ``characteristic width''. 
In this section, we attempt to 
understand why the distribution of widths is found to have such a narrow spread.

We follow the analysis of section \ref{sec:methods} on the \textit{Herschel} SPIRE-250 $\mu$m map of each cloud. 
We construct the distribution of widths measured at every cross-section (profile) of the filaments in the map. 
Studies finding a ``characteristic width'' have used the width of the mean profile of individual filaments to 
create the distribution of (mean filament) widths \citep[e.g. by fitting a Gaussian to the mean filament profile,][]{arzoumanian2011}. 
For comparison, we also construct the distribution of mean filament widths (as explained in section \ref{sec:methods}).

In Fig. \ref{fig:all_distr} we show the normalized distribution of filament-averaged widths
(dotted) and that of all profile widths (solid), for each of the clouds mentioned above. 
The number of filaments used to create these distributions as well as the mean and spread of the distributions are shown 
in Table \ref{tab:Nfils}. The initial fitting range used for all distributions was $\pm$0.1 pc from the filament ridge.
We find that the mean and spread of the distribution of mean filament widths for the filaments in IC 5146, 
and for those in all three clouds combined are in agreement with those found by \cite{arzoumanian2011} 
(the reported mean and spread were 0.1 pc and 0.03 pc, respectively).

When comparing the distribution of all filament profiles to that of mean widths we find that 
\textit{the shapes of the two kinds of distribution are clearly different.}
The filament-averaged width distribution is much more concentrated around its mean value 
($\sigma_{all} \approx [2-3] \times \sigma_{mean}$, from Table \ref{tab:Nfils}), 
and lacks the tails seen in the width distribution of all profiles.
The same effect is seen when filaments from all clouds are combined in a single distribution (bottom panel, Fig. \ref{fig:all_distr}). 

The differences between the two kinds of distributions can be easily understood considering the Central Limit Theorem (CLT).
The average value of a sample of profile widths (filament) is expected to follow a Gaussian distribution,
provided there is a sufficient number of filaments and that widths are not strongly correlated within a filament. 
This distribution of averages is centred around the mean of the parent distribution 
(that containing the widths of all filament profiles). 
It is therefore not surprising that the spread in the 
distribution of filament-averaged widths is small.
The information conveyed by the narrowness of this distribution is simply that the 
\textit{mean width of filaments is known with very good accuracy; not that the widths of individual filaments 
are constant}, as has often been interpreted.

We now wish to understand the factors that determine the spread of the distribution of mean filament widths 
(i.e. we wish to find the parameters $P$ that enter in $\sigma_{mean} = f(P)$). 
From the original form of the CLT (which assumes measurements are completely uncorrelated), we expect that 
\begin{equation}
\sigma_{mean} = \sigma_{all}/\sqrt{N}, 
\label{eqn:clt}
\end{equation}
if all filaments have the same number of profiles $N$ (which is proportional to the filament length). However, the lengths of filaments
follow a distribution of values $g(N)$, and therefore we expect $\sigma_{mean} = f(\sigma_{all}, g(N))$. 

In order to test whether these two parameters are sufficient to explain the observed $\sigma_{mean}$ in the Polaris Flare, 
we attempt to reproduce the distribution of mean filament widths from that of all profile widths as follows. 
We perform a Monte Carlo simulation where samples of widths are drawn randomly from the
observed distribution of all profile widths. These samples are randomly assigned to 100 groups (or fake `filaments'), 
corresponding to the 100 filaments found in the Polaris Flare. 
The size of each group
(corresponding to the length of the `filament') is drawn from the observed distribution of filament lengths. We then calculate the average
width of each group and construct the distribution of group-averaged widths. This process produces a distribution
with $\sigma^{rand}_{mean} = 0.004$ pc, much narrower than what is observed (see grey dashed distribution at the top panel of Fig. \ref{fig:all_distr}). 

The information that is lacking is that widths within the telescope beam size are strongly
correlated. We provide evidence for this by constructing the Autocorrelation Function of widths along the ridge of filaments, 
in appendix \ref{sec:appendix}. By performing a simulation similar to that
described above, including this final piece of information, we are able to reproduce the observed spread of the distribution of mean widths 
($\sigma_{mean} = 0.011$ pc, see appendix \ref{sec:appendix}). 
Therefore, the parameters that most significantly affect $\sigma_{mean}$ can be summarized as:
$\sigma_{mean} = f(\sigma_{all}, g(N), beam)$. In other words, \textit{there is no other information to be extracted from the spread of the filament-averaged width distribution.}

%First, because of this weak dependence, the CLT still applies, be it in a more general form \citep[e.g.][]{hilhorst}. If a filament
%had widths that were strongly dependent on the position along its spine, the CLT would not be valid and the distribution of filament-averaged widths
%would not necessarily be narrow. 
% http://sbfisica.org.br/bjp/files/v39_371.pdf
% http://people.math.umass.edu/~rsellis/pdf-files/limit-theorems-ellis-newman.pdf

Since the structure of individual filaments cannot be inferred uniquely from their average properties, we must 
examine how much does the width throughout an entire filament vary.
This question can be answered by considering the standard deviation of all widths measured
at different positions along the spine of a filament. 
Individual filaments are known to exhibit a range of FWHM along their spines
\citep[$0.1 -1$ pc, $0.07- 0.2$ pc, $0.1 -2.5$ pc][]{juvela2012a,malinen2012,andre2016}. 
In the Polaris Flare, the standard deviation of profile widths in a given filament is on average
$\left\langle \sigma_{fil} \right\rangle= 0.04$ pc.
This value is similar to the spread of the parent width distribution ($\sigma_{all} = 0.05$ pc, Table \ref{tab:Nfils}).
\textit{Therefore, the width varies significantly throughout the entire extent (length) of an individual filament.}
This finding is also supported by the two-dimensional distribution of filament widths across the Polaris Flare map, presented in
\cite{panopoulou2016} $-$ their figure 7. Filament widths do not exhibit large-scale regularities, but rather fluctuate 
in a seemingly random manner throughout the cloud.

\subsection{Why is the peak of the distribution of filament widths at 0.1 pc?}
\label{sec:explanation}

\begin{figure}
\centering
\includegraphics[scale=1]{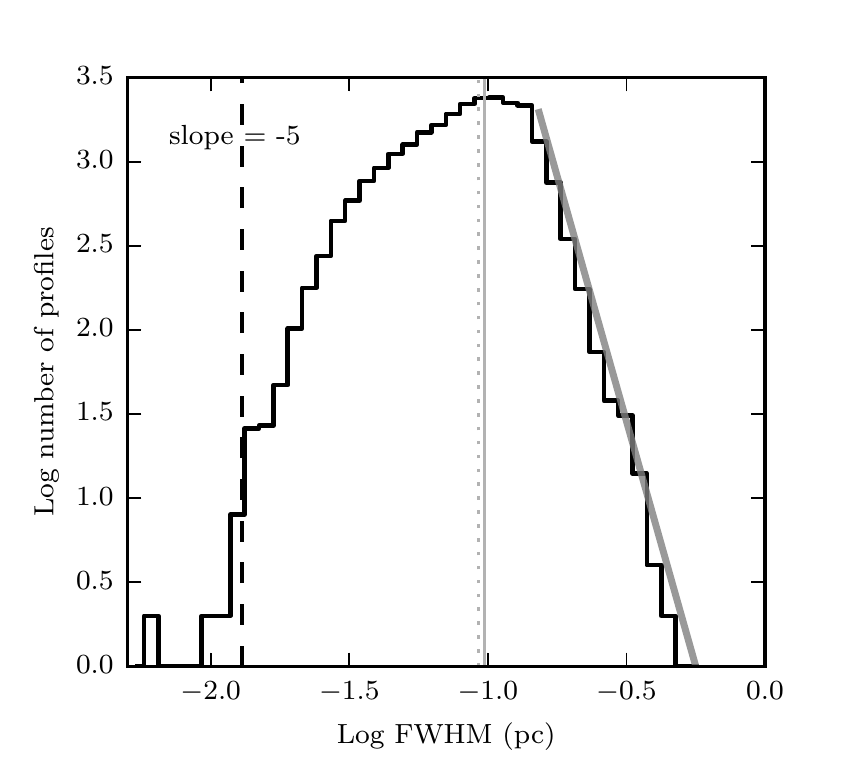}
\caption{Distribution of the logarithm of all profile widths of filaments in the Polaris Flare. 
At scales larger than 0.15 pc, the distribution resembles a power law 
(the grey solid line is a linear fit in log-log space to the logarithmically-spaced bins). 
The mean and median of the distribution are shown by the vertical solid and dotted lines, respectively. 
The \textit{Herschel}-SPIRE beam size (at 250 $\mu$m) at the distance of the Polaris Flare (150 pc) is shown with the dashed vertical line.}
\label{fig:pf-distr}
\end{figure}

The findings of section \ref{sec:distros} along with the scale-free power spectrum of the Polaris Flare, 
render the existence of a ``characteristic'' width of filaments highly unlikely. 
Consequently, the previously reported peak of the filament-averaged width distribution at 0.1 pc
cannot be explained by such a view.

In order to understand the origin and position of this peak, we examine in detail the width distribution of all filament profiles in 
the Polaris Flare. Fig. \ref{fig:pf-distr} shows the distribution of log(width) with equally-sized bins in logarithmic space.
It is strongly peaked and the mean and median values (vertical solid and dotted lines, respectively) differ by only $5\times10^{-3}$ pc. 
Interestingly, the part of the distribution at scales larger than the mean resembles a straight line (in logarithmic space). 
A linear fit (in log-log space) to the distribution is shown with a grey solid line in Fig. \ref{fig:pf-distr}.

This resemblance of part of the distribution to a power-law is not surprising if we consider the evidence for scale-free, 
or self-similar, structures in the ISM within a range of scales \citep[e.g.][]{stutzki1998, elmegreen2004, tassis2007, elia2014, miville2016}. \cite{elmegreenfalgarone} find size and mass distributions in clouds consistent with those arising from a 
fractal. In the case of the Polaris Flare, the scale-free nature of the cloud is evidenced by its power-law spatial power spectrum
\citep{miville}.
We note that because the Polaris Flare is gravitationally unbound and is not forming stars \citep[][]{heithausen2002, ward-thompson, wagle}, 
self-similarity is not expected to break down at the typical scale of prestellar cores \citep[0.1 pc, e.g.][]{goodman1998}.
Therefore, a power-law distribution of filament widths (in accordance to other length scales) within some range of scales,
is not unreasonable. 

%Second, the observed distribution of widths is merely a probe
%of the true, intrinsic distribution of structure sizes in the cloud, within the range of scales covered by the \textit{Herschel} map. 

If a power law size (width) distribution is intrinsic to the cloud, we expect that this power law will be truncated at large scales 
at a fraction of the size of the \textit{Herschel} map and at small scales (at least) by the resolution. 
In the process of imaging the cloud dust emission and measuring sizes on the map, errors are introduced. 
Errors have the effect of smoothing the distribution near the lower truncation limit \citep[][]{koenkondlo}. 
If these errors are assumed Gaussian, then the shape of the (truncated) power law distribution can be analytically modelled 
\citep[][]{koenkondlo} and is similar to that of the distribution in Fig. \ref{fig:pf-distr}: 
it possesses a peak near the lower truncation limit followed by a power law tail.
The analytical form of the distribution of measured widths ($W$) is:
\begin{equation}	
f(W) = \int^{W_{max}}_{W_{min}} \frac{\gamma w^{-(\gamma+1)}}{\sqrt{2\pi}\sigma(W^{-\gamma}_{min} - W^{-\gamma}_{max})}
			\exp\left[ -\frac{1}{2} (\frac{W - w}{\sigma})^2 \right] dw,
\label{eqn:powerlaw}	
\end{equation}
%Elmegreen \& Falgarone (1996) pointed out that the size and mass distribution of clouds are consistent with originating from a fractal 
%structure across scales that range from 0.01 to 100 pc. Size distributions were shown to follow power laws in the range of several
%times the resolution size, as expected for the size distribution of substructures in a fractal (mandelbrot 1983).
%In length scales from hundreds of parsecs down to 0.01 pc the scale-free nature of the ISM is manifested in the featureless power spectra 
%of clouds such as the Polaris Flare (Bensch 2001, miville) and high galactic latitude cirrus (miville 2003, miville 2016) in the clouds of 
%the Herschel Hi-Gal survey (elia et al) and in HI maps (e.g. miville2003). 
where $w$ is the width before introducing measurement errors, $\sigma$ is the measurement uncertainty,
$\gamma +1 $ is the power law slope, 
and $W_{min}, W_{max}$ are the sizes at which the power 
law is truncated due to (at least) the resolution and (at most) the map size. 
For what follows, we will consider only non-negative values of $\gamma$. 
The average value can be obtained by (numerical) integration of the formula:
\begin{equation}
\langle W \rangle = \int^{\rm map \, size}_{\rm HPBW}{f(W)W}dW,
\label{eqn:expectation}
\end{equation}
where the integration is performed within the bounds set by the observations (resolution limit and map size).

\begin{figure}
\centering
\includegraphics[scale=1]{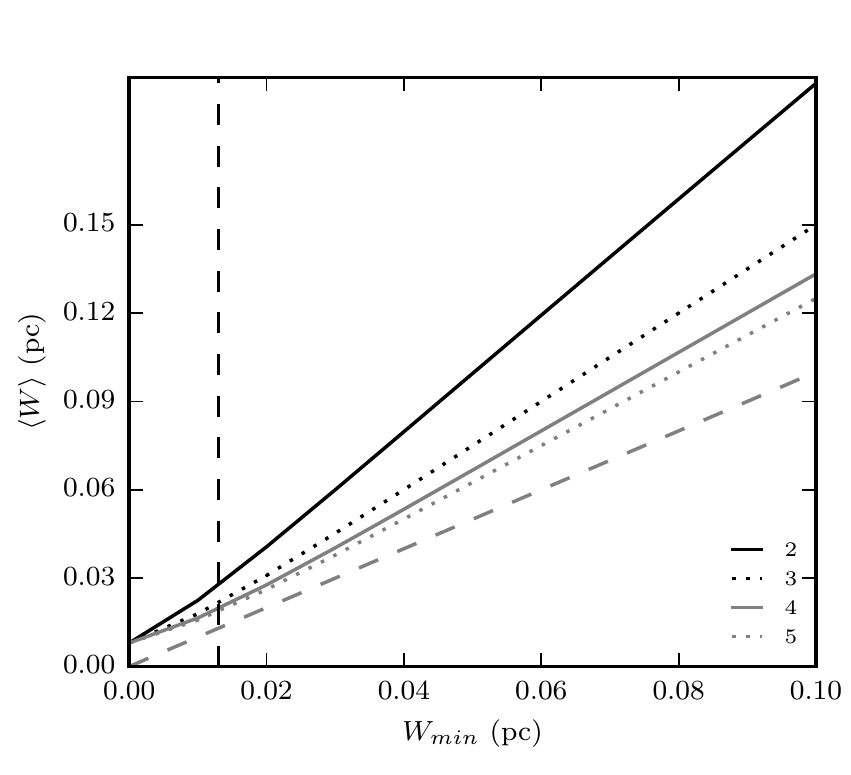}
\caption{Dependence of the mean measured width of profiles on $W_{min}$ for different values of $\gamma$ (ranging from 2 to 5 and marked with labelled lines).
Values of $W_{max} = 5$ pc and $\sigma = 0.02$ pc were used, since there is little dependence 
of $\langle W \rangle$ on these parameters (see text). A 1$-$1 correlation is shown with the dashed grey line. 
The dashed vertical line shows the \textit{Herschel}-SPIRE beam size (at 250 $\mu$m) in parsecs at the distance
to the Polaris Flare (150 pc).}
\label{fig:wmean-dependence}
\end{figure}

Therefore, if one could determine the parameters $\sigma$, $\gamma, W_{min}$, and $W_{max}$, a prediction for the mean of 
the distribution of all profile widths could be obtained. In section \ref{sec:distros} we found that \textit{the mean of 
the distribution of all profile widths is coincident with the position of the peak of the filament-averaged distribution (as a result
of the CLT). Consequently, measuring the mean of the former kind of distribution determines the peak of the latter.} We note
that the mean and peak of the distribution of filament-averaged widths are the same, as it is (approximately) Gaussian.

In Fig. \ref{fig:wmean-dependence} we explore the dependence of $\left\langle W\right\rangle$ 
on the parameters $W_{min}$ and $\gamma$. $\langle W \rangle$ is plotted against $W_{min}$ for $W_{max} = 5$ pc,
$\sigma = 0.02$ pc, and different values of $\gamma$ (ranging from 2$-$5, 
around a value of 4 implied by the slope of Fig. \ref{fig:pf-distr}, and marked with solid and dotted lines). 
%The black horizontal line is the mean of the width distribution found in \cite{arzoumanian2011} and the grey band 
%corresponds to the range within $\pm 1 \sigma$ of their distribution. 
%For $W_{min}$ in the range $0.05 - 0.09$ pc equation \ref{eqn:expectation2} takes values within the
%spread of the distribution found in \cite{arzoumanian2011}.
We find that there is no dependence of $\langle W \rangle$ on $W_{max}$. 
For $\gamma$ in the range 2$-$5, constant $W_{min}$ and $\sigma$, and $W_{max}$ in the range $1-50$ pc,
$\langle W \rangle$ varies by less than 0.01 pc. Similarly, for constant $W_{min}$ and $W_{max}$, 
and $\sigma$ in the range 0.01$-$0.1 pc, $\langle W \rangle$ varies by less than 0.02 pc for any given $\gamma$ within 2$-$5.
On the contrary, $\left\langle W\right\rangle$ is very sensitive to the parameter $W_{min}$. 
\textit{Therefore, the lower scale at which a power law width distribution is truncated essentially sets the position of the 
peak in the distribution of filament-averaged widths.}

\begin{figure}
\centering
\includegraphics[scale=1]{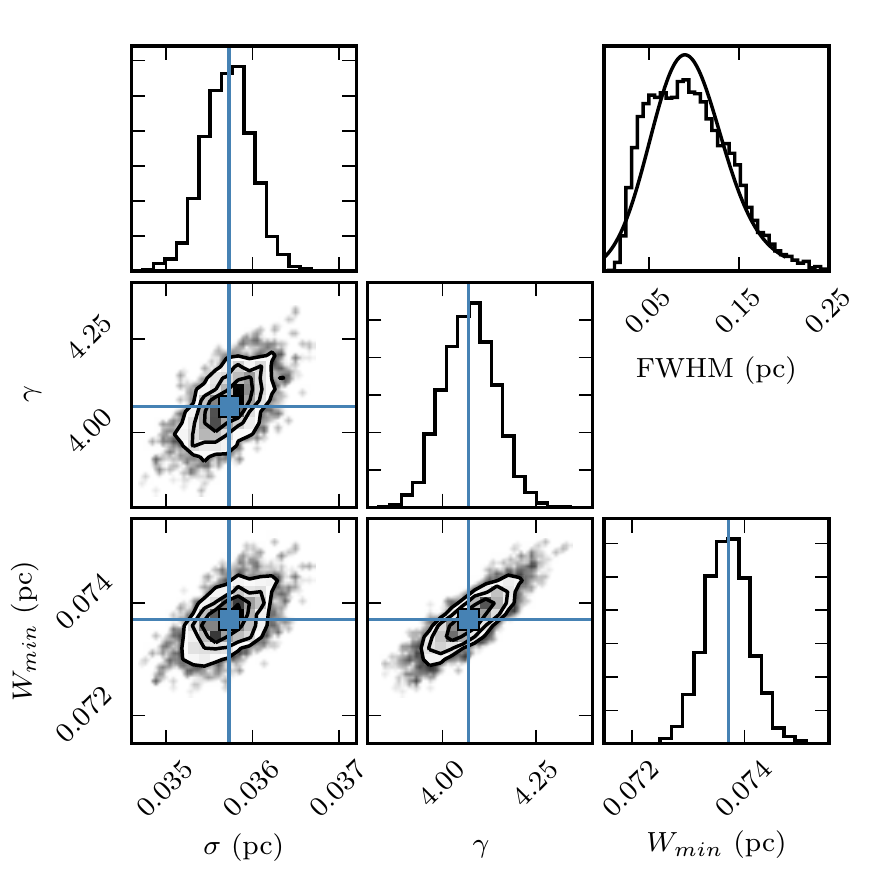}
\caption{The posterior distribution of parameters from our MCMC model of the data of Fig. \ref{fig:pf-distr}. 
Blue lines show the values returned by the MLE.
On diagonal: One-dimensional histograms of model parameters: $\sigma$, $\gamma$, and $W_{min}$.
Lower diagonal: Joint PDFs of the posterior distribution of the model parameters.
Top right: comparison between normalized distribution of Fig. \ref{fig:pf-distr} (stepped histogram)
with our model (equation \ref{eqn:powerlaw}), using a randomly selected set of parameter values
from the posterior distribution of model parameters (smooth line).}
\label{fig:MCMC}
\end{figure}

We would like to identify the best fit parameters for the distribution of filament widths in the Polaris Flare, 
using a power law with measurement uncertainties. \cite{koenkondlo} provide the log-likelihood function for this model:
\begin{equation}
\begin{split}
\mathcal{L} = -\frac{N}{2} \log 2\pi -N \log \sigma + N \log \gamma - N \log(W_{min}^{-\gamma} - W_{max}^{-\gamma})\\
+ \sum_{j=1}^N \log \int_{W_{min}}^{W_{max}} x^{-(\gamma +1)} \exp\left[-\frac{1}{2} \left(\frac{W_j -x}{\sigma} \right)^2 \right] dx
\end{split}
\label{eqn:likelihood}
\end{equation}
where $W_j$ are the $N$ different filament width measurements and $\sigma$ is the Gaussian standard deviation on measurements of $W_j$
(and is independent from $W_j$). 
Since, for the distributions of filament widths, the effect of $W_{max}$ is insignificant, we set it to 10 pc 
(approximate size of the \textit{Herschel} map) 
and solve for the best-fit values of $W_{min}$, $\gamma$, and the measurement uncertainty, $\sigma$. 
We use the routine {\tt minimize} within the {\tt scipy.optimize} package, to find the values that maximize the log-likelihood.  
The resulting values for the parameters are:
$\sigma = 0.036$ pc,  $\gamma = 4.07$, and $W_{min} = 0.074$ pc. 
We note that the slope found by fitting a line to the distribution in Fig. \ref{fig:pf-distr} (slope = 5) is in good
agreement with the value found by maximizing the log-likelihood ($\gamma$ + 1 = 5.07).
By substituting $\gamma$ and $W_{min}$ in equation \ref{eqn:expectation}, 
we obtain $\left\langle W\right\rangle = $ 0.098 pc, which is approximately equal to the mean of the observed distribution of widths.

Additionally, we would like to determine the range of parameter values implied by the filament width data. 
Instead of using a maximum likelihood estimate (MLE), we use the Markov Chain Monte Carlo (MCMC) algorithm {\tt emcee} \citep{foreman-mackey2013}. 
{\tt emcee} employs an affine-invariant ensemble sampler to probe the model parameter space. Our model uses 32 walkers to maximize the 
log-likelihood function in Equation \ref{eqn:likelihood}. We apply flat priors on $W_{min}$ and $\gamma$, in the ranges:
$W_{min}$: [0$-$0.5] pc, $\gamma$: [0.01$-$10]. We use the Jeffreys prior ($1/ \sigma$) on $\sigma$ to make it scale-invariant. 
The range used for $\sigma$ is [0$-$3] pc. We note that $\sigma$ encapsulates the uncertainty introduced by two different processes: 
the imaging of the cloud (resolution) and the measurement of the width (analysis in section \ref{sec:methods}).
We therefore choose to leave $\sigma$ as a free parameter (and do not set it equal to the image resolution) 
to account for both sources of error.
200 steps are sufficient for the ``burn-in'' stage. We throw these data away, and run our model for 2000 additional steps 
to produce the posterior distribution.

The posterior distributions of model parameters (Fig. \ref{fig:MCMC} $-$ on diagonal) are strongly peaked, with 
standard deviations of $3 \times 10^{-4}$ pc for $\sigma$, 0.07 for $\gamma$, and $4 \times 10^{-4}$ pc for $W_{min}$.
Joint PDFs of the posterior distributions of model parameters are shown in the panels lower than the diagonal. 
As expected for uniform priors (and since our prior on $\sigma$ is weak), 
the region of high probability parameter space agrees with the results from the maximum likelihood estimate (blue lines).
The top right panel in Fig. \ref{fig:MCMC} compares the distribution of filament widths from Fig. \ref{fig:pf-distr}
(histogram) to the functional form of equation \ref{eqn:powerlaw} with values for the parameters drawn randomly from
the posterior distribution of model parameters (smooth line). 
For the range of parameters used, the shape of the smooth curve varies very little 
(the variation is similar to the width of the plotted line).
The model captures well the basic shape of the distribution and of the model parameter space.

For values of the parameters within 5$\sigma$ of the mean of their corresponding distributions,  
we obtain from equation \ref{eqn:expectation}: $\langle W \rangle = 0.09-0.1$ pc. We conclude that the model used here 
accurately reproduces the position of the peak of the distribution of widths of Fig. \ref{fig:pf-distr} (within 0.01 pc).
From the value of $\sigma$ we obtain a handle on the error introduced by the width calculation algorithm. 
Since the resolution is only $0.013$ pc, the algorithm is the main source of measurement error.

The question that remains to be answered is what determines $W_{min}$, the value below which the power law distribution is truncated.
One obvious culprit could be the resolution limit. However, if $W_{min}$ were equal to the telescope beam size (0.013 pc, 
shown by the vertical dashed line in Fig. \ref{fig:wmean-dependence}), 
the mean of the distribution would fall below 0.03 pc, as seen from Fig. \ref{fig:wmean-dependence}.
Another possibility would be that the combined effect of the telescope resolution and the errors of the width-measurement algorithm
are setting the lower limit. This corresponds to the parameter $\sigma$ which is almost 3 times the beam size 
(for the distribution of widths in Fig. \ref{fig:pf-distr}). 
$W_{min}$, however, is found to be almost 6 times the beam size, making both options unlikely.

\begin{table}
\centering
\caption{Values of the parameters returned by the MLE for distributions with different initial fitting ranges. 
The final column shows the values derived from the parameters using equation \ref{eqn:expectation}.}
\begin{tabular}{|c|c|c|c|c|}
\hline
Fitting range & $\sigma$  & $\gamma$ & $W_{min}$  & $\left \langle W \right \rangle$ \\
(pc)          &   (pc)    &         & (pc)        & (pc)  \\
\hline
\hline
0.04 & 0.012 & 4.65 & 0.037 & 0.05\\
\hline
0.06  & 0.019 & 3.97 & 0.049 & 0.07   \\
\hline
0.08 & 0.027 & 4.13 & 0.064  & 0.08 \\
\hline
0.10		& 0.036 & 4.07 & 0.074  & 0.10 \\
\hline
0.12 	& 0.043 & 4.25 & 0.087  & 0.11\\
\hline
0.14	 & 0.051 & 4.16 &  0.094  & 0.12\\
\hline
0.20 & 0.065 & 3.24 & 0.110   & 0.16\\
\hline
0.25 & 0.079 & 3.03 & 0.125 & 0.19\\
\hline
%\footnotesize{$^*$ $\rm s_{los} = N_H/n_H$.}
\end{tabular}
\label{tab:MLE}
\end{table}

\begin{figure}
\centering
\includegraphics[scale=1]{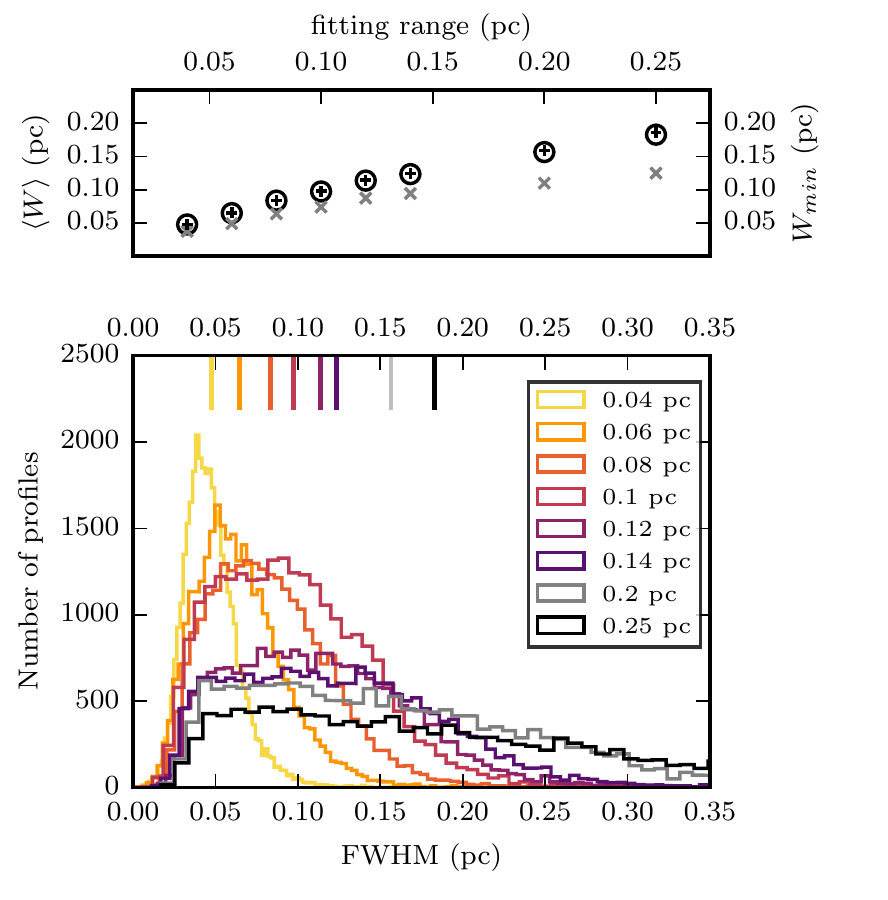}
\caption{Top: the mean of the observed distributions (open circles) 
and that calculated from equation \ref{eqn:expectation} (crosses) with the MLE values of the parameters, versus the fitting range (left vertical axis). Values of $W_{min}$ returned by the MLE for each distribution as a function of fitting range
are shown as grey crosses (right vertical axis).
Bottom: Distributions of beam-deconvolved profile widths in the Polaris Flare for different values of initial fitting range 
(distance from the axis of a filament). 
The vertical lines at the top show the mean value of each distribution.}
\label{fig:range}
\end{figure}

A more likely possibility is that $W_{min}$ is related to the range over which the Gaussian fit is performed to measure the width 
of a profile. In order to avoid fitting the wings of filament profiles, studies of filament widths have chosen to fit a Gaussian 
within a range of $\sim$ 0.1 to 0.4 pc from the filament spine \citep{arzoumanian2011, juvela2012a, smith2014, koch2015, andre2016, federrath2016}.
%Keep an eye out for Summeye Suri's paper on CO filaments in Orion. 0.04 pc?
%http://arxiv.org/pdf/1603.09177v1.pdf}. 
However, \cite{smith2014} have shown that the selection of fitting range drastically affects the mean filament width. 

We investigate this effect further for the filaments in the Polaris Flare, by creating distributions of widths with
different initial fitting ranges, and repeating the MLE analysis for each one. The resulting values of the parameters
are listed in Table \ref{tab:MLE}. 
As can be seen in Fig. \ref{fig:range}, the peak of the distribution of widths shifts towards lower values for smaller fitting ranges 
(vertical line segments on the top of the main panel). 
From the top panel of Fig. \ref{fig:range} (left vertical axis), we find that the mean of the observed distributions 
(open circles) is in very good agreement with the mean calculated with equation \ref{eqn:expectation} using the values of 
Table \ref{tab:MLE} (black crosses).
Both quantities increase monotonically with the fitting range.
This is because $W_{min}$ also has such a dependence on the fitting range (Fig. \ref{fig:range} top panel, 
right vertical axis). 
\textit{Since the fitting range used in previous studies does not vary much, it is not surprising that the 
peak of the filament-averaged width distribution is found at similar values for different clouds.}

Compared to the simulations of \cite{smith2014}, 
the mean width of filaments in the Polaris Flare 
increases slightly more abruptly as a function of fitting range. For a factor of $\sim$ 3 increase in fitting range, these authors
find the mean width to increase by a factor of 1.5, whereas we find a factor of $\sim$ 2. 
However, any such comparison must consider the range that we use as an upper limit, because we fit Gaussians 
iteratively beginning from the quoted fitting range. Also, there is an indication in Fig. \ref{fig:range} that at larger ranges the 
slope tends to flatten out. Considering these factors, we find the scaling of mean width with fitting range to be strikingly similar
between the simulated and observed filaments, perhaps hinting at common structural properties.

The dependence of the mean width on the fitting range suggests that selecting a scale over which to 
observe/measure a structure determines how important the inner-most part of the structure will appear\footnote{Observational 
evidence for substructure in \textit{Herschel} filaments (through finer resolution imaging) already exists, 
see e.g. \cite{fernandez2014, henshaw2016a, henshaw2016b}.}.
In the case of a scale-free structure this can be understood, as a fractal surface 
changes in character when examined at different scales: it appears smoother/flatter when observed from further away, but upon
closer inspection substructure appears. Alternatively, for a structure with a well-defined peak, a Gaussian fit with offset will
always return a narrower FWHM for a smaller fitting range.

\section{SUMMARY}
\label{sec:summary}

In this work we have explored the analysis of filament widths in an attempt to find a way to reconcile the proposed
``characteristic'' width of filaments with the absence of its imprint in spatial power spectra. Our findings can be summarized in the 
following points:
\begin{itemize}
\item[i.] The selected methodology for measuring widths can produce a peaked distribution even if the original data do not contain a 
preferred scale.
\item[ii.] The process of averaging over filament profiles results in a distribution that is necessarily narrow, as a result of the 
central limit theorem.
\item[iii.] Widths vary significantly as a function of position on the spine of a filament.
\item[iv.] The position of the previously identified peak (0.1 pc) in \textit{Herschel} data could be determined by the choice of distance from the 
filament spine within which the width is measured.
\end{itemize}
The above suggests that filaments are unlikely to have a constant width, a result which explains the lack of a characteristic scale
in the spatial power spectrum of the Polaris Flare. 
%Furthermore, the power-law power spectrum of the Polaris Flare, and the dependence of 
%the filament width on the scale over which it is measured point to a view in which filaments are self-similar
%structures. This idea is also supported by observations of substructure inside \textit{Herschel} filaments:
%\citep[e.g.][]{henshaw2016a, henshaw2016b}. 
%Also, under this view it is expected that when filaments are viewed at larger distances their substructure will be smoothed out, and so they 
%would appear thicker. This could explain the findings of \cite{schisano} and \cite{rivera-ingraham} that filaments in Hi-Gal maps become 
%wider with distance. 
Finally, we note that our results are specific for the widths of filaments and do not contradict
the (well-established) existence of other typical length scales in clouds \citep{mouschovias1991}.

\section*{Acknowledgements}

We would like to thank Dmitriy Blinov, Vasiliki Pavlidou, and Aris Tritsis for fruitful discussions as well as Vassilis Charmandaris, 
Paul Goldsmith, Nick Kylafis, and Josh Peek for their helpful comments on the paper. 
We are grateful to the anonymous reviewer for a detailed report which greatly improved this work.
Also, we thank Monica He for her contribution during the first steps of this work and Damianos Mylonakis for helping out 
with technical issues. We used the python module {\tt corner} \citep{foreman-mackey-triangle} to produce Fig. \ref{fig:MCMC}. 
This research made use of Astropy, a community-developed core Python package for Astronomy \citep{astropy}.
G.V.P. and K.T. acknowledge support by FP7 through the Marie Curie Career Integration Grant 
PCIG-GA-2011-293531 ``SFOnset'' and partial support from the EU FP7 Grant PIRSES-GA-2012-31578 ``EuroCal''.
J. J. A. acknowledges funding from the European Research
Council under the European Union's Seventh Framework Programme (FP/2007-2013)/ERC Grant
Agreement n. 617001.

This research has used data from the \textit{Herschel} Gould Belt Survey (HGBS) project 
({\tt http://gouldbelt-herschel.cea.fr}). The HGBS is a \textit{Herschel} Key Programme jointly carried out by SPIRE Specialist Astronomy 
Group 3 (SAG 3), scientists of several institutes in the PACS Consortium (CEA Saclay, INAF-IFSI Rome and INAF-
Arcetri, KU Leuven, MPIA Heidelberg), and scientists of the \textit{Herschel} Science Center (HSC).

%%%%%%%%%%%%%%%%%%%%%%%%%%%%%%%%%%%%%%%%%%%%%%%%%%



%\bibliographystyle{mnras}


%%%%%%%%%%%%%%%%%%%%%%%%%%%%%%%%%%%%%%%%%%%%%%%%%%

%%%%%%%%%%%%%%%%% APPENDICES %%%%%%%%%%%%%%%%%%%%%

\appendix

\section{{\tt DISPERSE} skeletons of Herschel images}
\label{sec:skeletons}

As explained in section \ref{sec:methods}, the first step in the analysis of filament widths is the identification of the 
filaments in the image.
This is done here by using {\tt DISPERSE} \citep[][]{sousbie2011} to acquire the skeleton of the image and then using 
{\tt FILTER} \citep{panopoulou2014} to post-process the skeleton and discard spurious structures.

\begin{figure}
\centering
\includegraphics[scale=1]{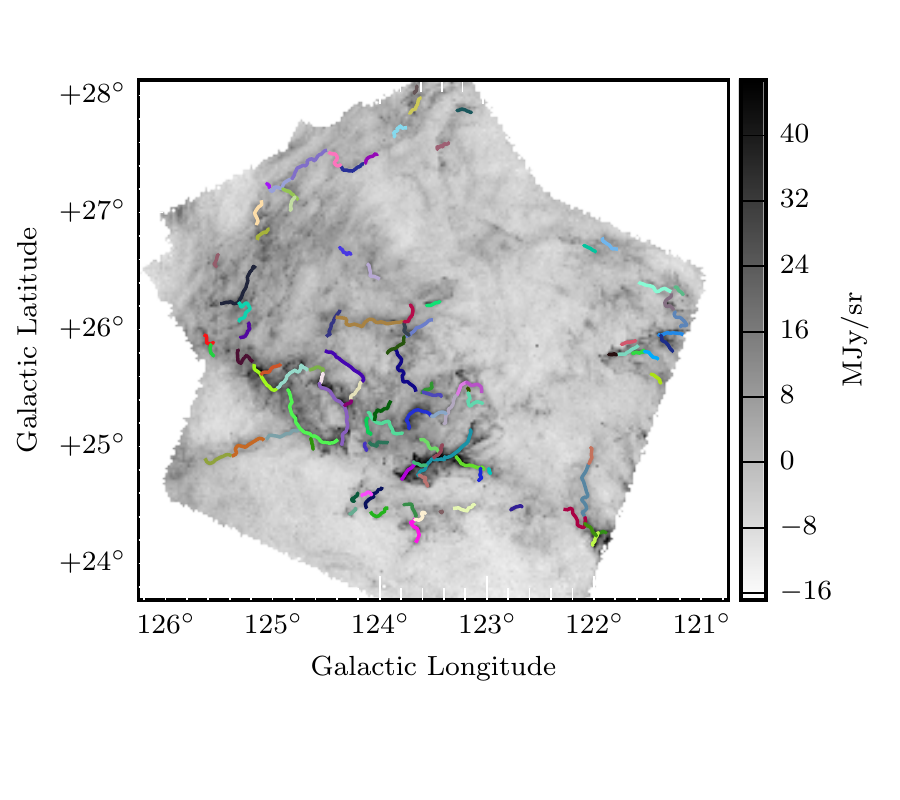}
\caption{Skeleton of the Polaris Flare 250 $\mu$m \textit{Herschel} image constructed with {\tt DISPERSE} and post-processed with {\tt FILTER}. Coloured lines trace the spines of filaments used in our analysis.}
\label{fig:pf}
\end{figure}

\begin{figure}
\centering
\includegraphics[scale=1]{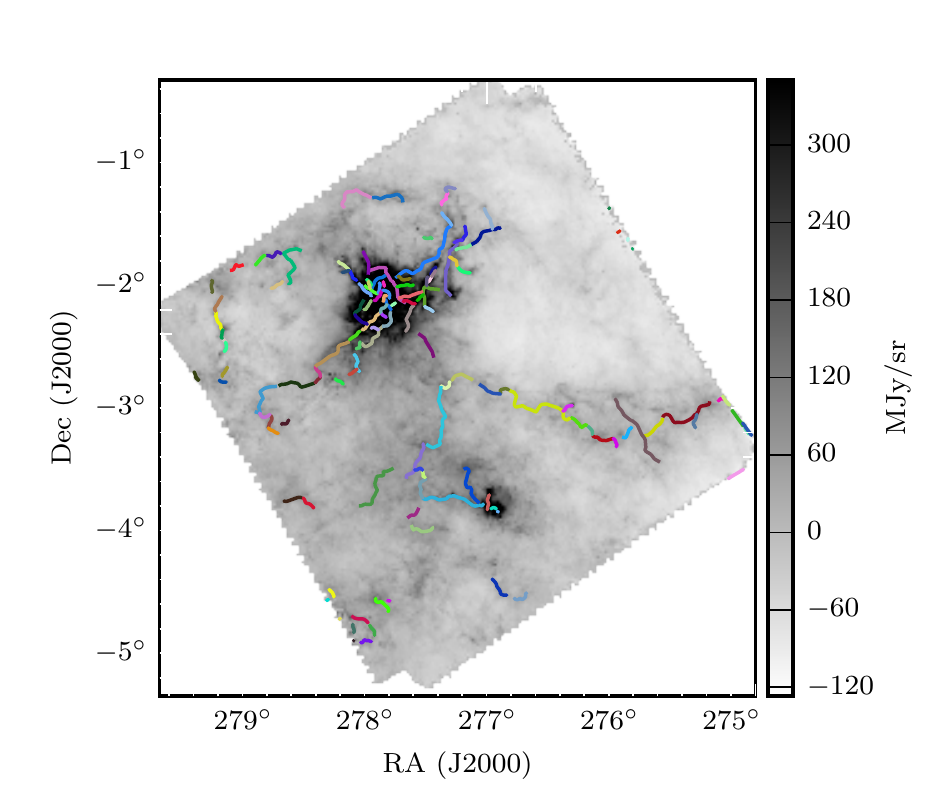}
\caption{Same as Fig. \ref{fig:pf} but for the image of Aquila.}
\label{fig:aq}
\end{figure}

\begin{figure}
\centering
\includegraphics[scale=1]{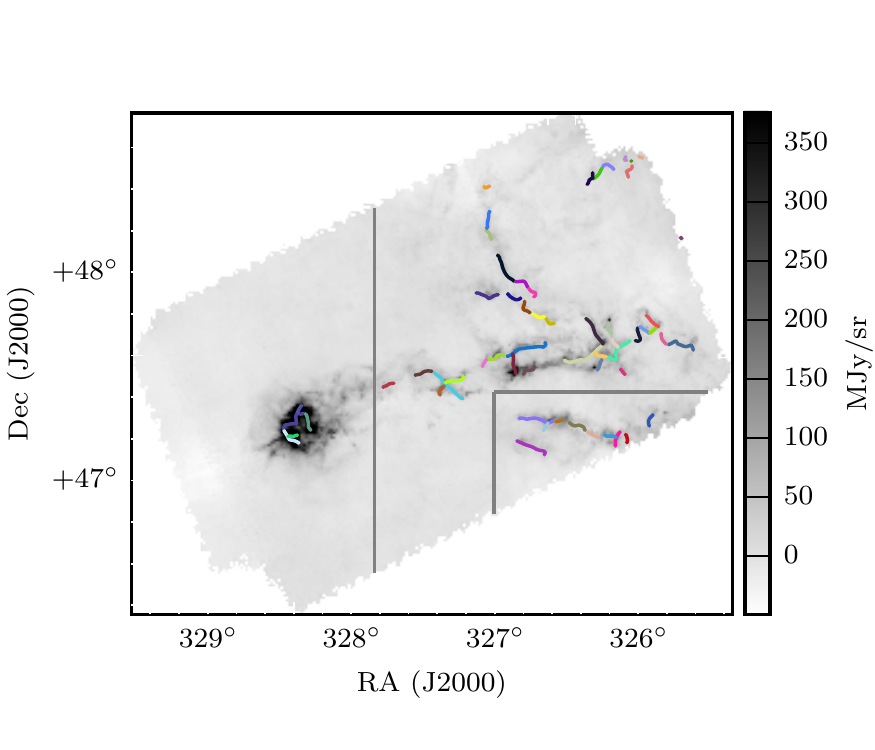}
\caption{Same as Fig. \ref{fig:pf} but for the image of IC 5146. The grey lines divide the image into three submaps on which 
{\tt DISPERSE} was run separately.}
\label{fig:ic}
\end{figure}

The (post-processed) skeletons of the three clouds (Polaris Flare, Aquila and IC 5146) used in our analysis are shown 
in Figs. \ref{fig:pf} $-$ \ref{fig:ic} (coloured lines) overplotted on the 250 $\mu$m \textit{Herschel} images. 
Each line represents a single structure (filament). The parameters of {\tt DISPERSE} for these skeletons are shown in table
\ref{tab:disperse}. The skeletons obtained are very similar to those of \cite{arzoumanian2011,konyves2015,andre2014}.

For the Polaris Flare and Aquila, {\tt DISPERSE} was run directly on the entire (unfiltered) \textit{Herschel} image. 
The skeleton for IC 5146 was produced by running {\tt DISPERSE} on three sub-maps (divided by grey lines in Fig. \ref{fig:ic}) 
and combining the resulting skeletons. 
This enabled us to isolate regions of similar intensity, as in the whole map the differences in brightness caused either faint 
structures not to be identified or spurious structures to be identified in the brightest parts. In table \ref{tab:disperse}, 
indices 1, 2, and 3 refer to the left, middle and bottom regions of the map respectively.

We have performed a parameter study for the skeletons of the image of Aquila to test for effects on
the distribution of filament profile widths. 
The ranges of parameters used \citep[for parameter definitions see][]{sousbie2011} were: persistence 60 $-$ 80 
MJy/sr, robustness 82 $-$ 102 MJy/sr, smoothing 50 $-$ 200, and assembling of arcs 50 $-$ 90 degrees.
All resulting distributions of profile widths were identical.

\begin{table}
\centering
\caption{Parameters of {\tt DISPERSE} used for the skeletons of Fig. \ref{fig:pf} $-$ Fig. \ref{fig:ic}.}
\begin{tabular}{|c|c|c|c|c|}
\hline
Cloud & persistence & robustness  & smooth  & assemble \\
&(MJy/sr) &(MJy/sr)&&(degrees)\\
\hline
\hline
Polaris Flare & 15 & 16 & 50 & 60\\
\hline
Aquila &  80 & 82 & 50 &  50  \\
\hline
IC 5146 1 & 400 & 410 & 60  & 90 \\
IC 5146 2 & 50 & 51 & 100  & 60 \\
IC 5146 3 & 40 & 55 & 100  & 40 \\
\hline
\end{tabular}
\label{tab:disperse}
\end{table}

\section{Can a characteristic scale be `hidden' from the power spectrum?} 
\label{sec:powerspectra}
In this section we use simple artificial images to investigate the effect of a characteristic scale on the 
azimuthally averaged spatial power spectrum of an image. 

\begin{figure*}
\centering
\includegraphics[scale=1]{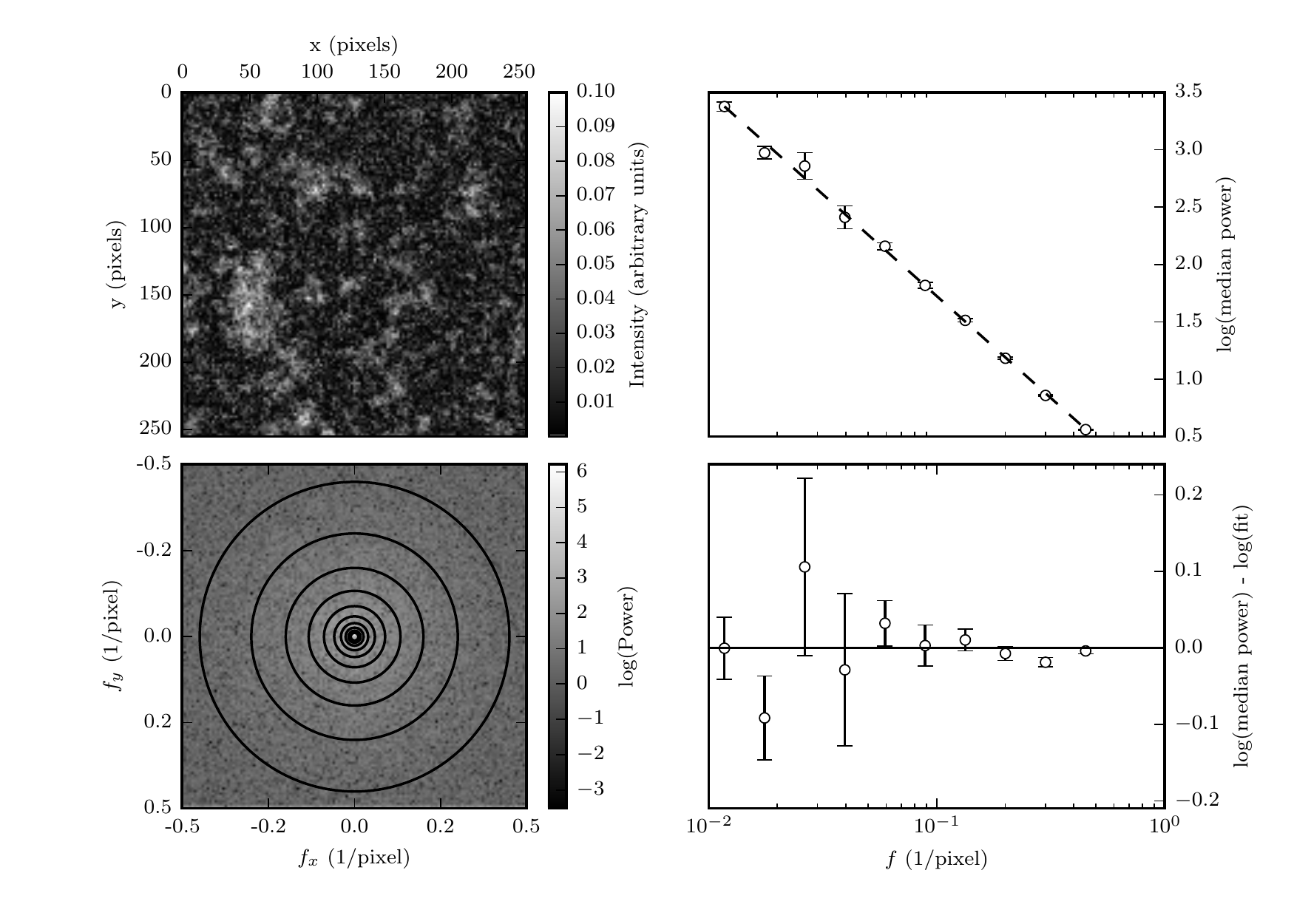}
\caption{Top left: two-dimensional pink noise that is used in subsequent images of artificial filaments. 
Bottom left: two-dimensional spatial power spectrum of the image with annuli drawn to indicate the regions within which
the azimuthally-averaged power spectrum (top right) is calculated. Top right: Azimuthally averaged spatial power spectrum 
with errors and linear fit performed in log-log space (dashed line).
Bottom right: Difference between the (logarithms of the) spatial power spectrum and the fit.}
\label{fig:pinknoise}
\end{figure*}

Our tests consist in creating elongated structures with radial profiles following the form of a Plummer profile
in column density \citep[a form which fits well the column density profiles of observed filaments,][]{arzoumanian2011}. 
The intensity of the profile of an artificial filament is $I_0$ on its spine, 
has an inner flat portion (in logarithmic space) of size $\rm R_{flat}$,  
and drops with distance ($r$) from the axis of the filament as:
\begin{equation}
\centering
 I(r) = \frac{I_0}{[1+(r/{\rm R_{flat}})^2]^{\frac{p-1}{2}}} .
\label{eqn:plummercolumn}
\end{equation}
For the exponent, $p$, we choose a value of $p=2$, as observed for filaments in \textit{Herschel} data \citep{arzoumanian2011}

\begin{figure*}
\centering
\includegraphics[scale=1]{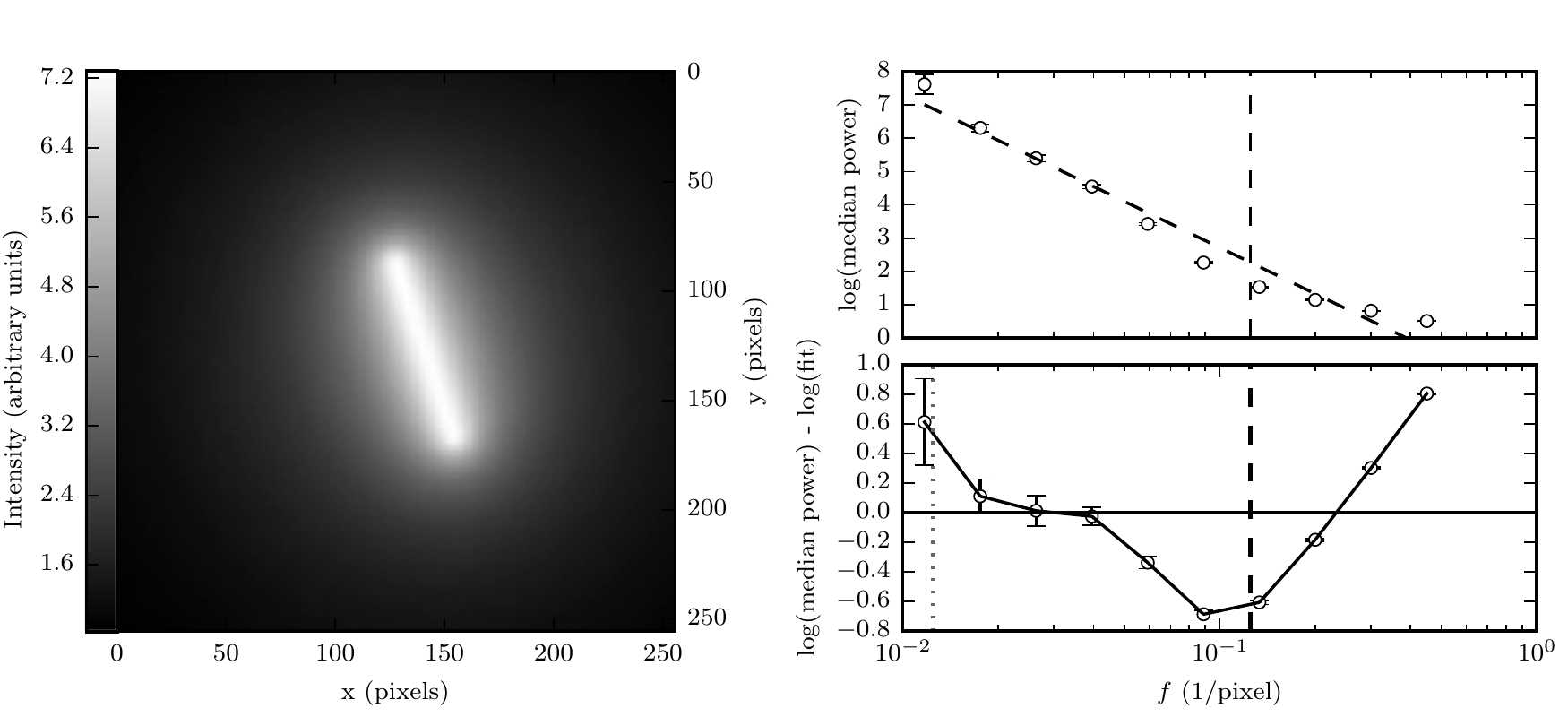}
\caption{Left: Artificial image of a single filament with $\rm R_{flat} = 8$ pixels and length = 80 pixels. 
Top right: spatial power spectrum of the image with a linear fit in log-log space (dashed line). 
Bottom right: difference between the (logarithms of the) power spectrum and fit. 
The dashed vertical line shows the spatial scale corresponding
to the width $\rm R_{flat}$, while the dotted line shows the scale corresponding to the length. 
Low intensity pink noise (Fig \ref{fig:pinknoise}) has been added to the image of the filament.}
\label{fig:onefilimg}
\end{figure*}

The artificial filaments have very simplistic characteristics: they are straight, have a constant peak intensity along their spine, 
and a constant $\rm R_{flat}$. This should make the identification of the signature of any characteristic scale (width, length), 
in the spatial power spectrum, unambiguous. 
In this section we use $\rm R_{flat}$ as a proxy of the width of the artificial filaments as 
for profiles with $p =2$, $\rm FWHM \approx 3 {\rm R_{flat}}$ \citep{arzoumanian2011}.
 
The filament images are co-added with (the same) two-dimensional isotropic pink noise.
We generate the noise by creating the coefficients of its two-dimensional Fourier transform. 
Each coefficient $C_k$ has a magnitude of $1/\sqrt{k^2_x + k^2_y}$ and a random phase 
($C_0$ is set to 0). We obtain the final pink noise image by applying the inverse Fourier transform (Fig. \ref{fig:pinknoise}, top left). 
Its two-dimensional power spectrum is shown in the bottom left panel. 
The azimuthally-averaged power spectrum \citep[constructed by taking the median power within the annuli drawn on the 
two-dimensional power spectrum, as in][]{pingel} has the form of a power-law (Fig. \ref{fig:pinknoise}, top right). 
The deviation from a perfect power-law can be quantified by the residuals of the power spectrum from a linear fit in log-log space
(dashed line in top right panel). From the bottom right panel of Fig. \ref{fig:pinknoise} we see that the level of the residuals,
log(median power) - log(fit), is less than 0.2. As in section \ref{sec:fractal}, the error of the median of each annulus is
quantified by bootstrap resampling the distribution of intensities within the annulus. 

\begin{figure}
\centering
\includegraphics[scale=1]{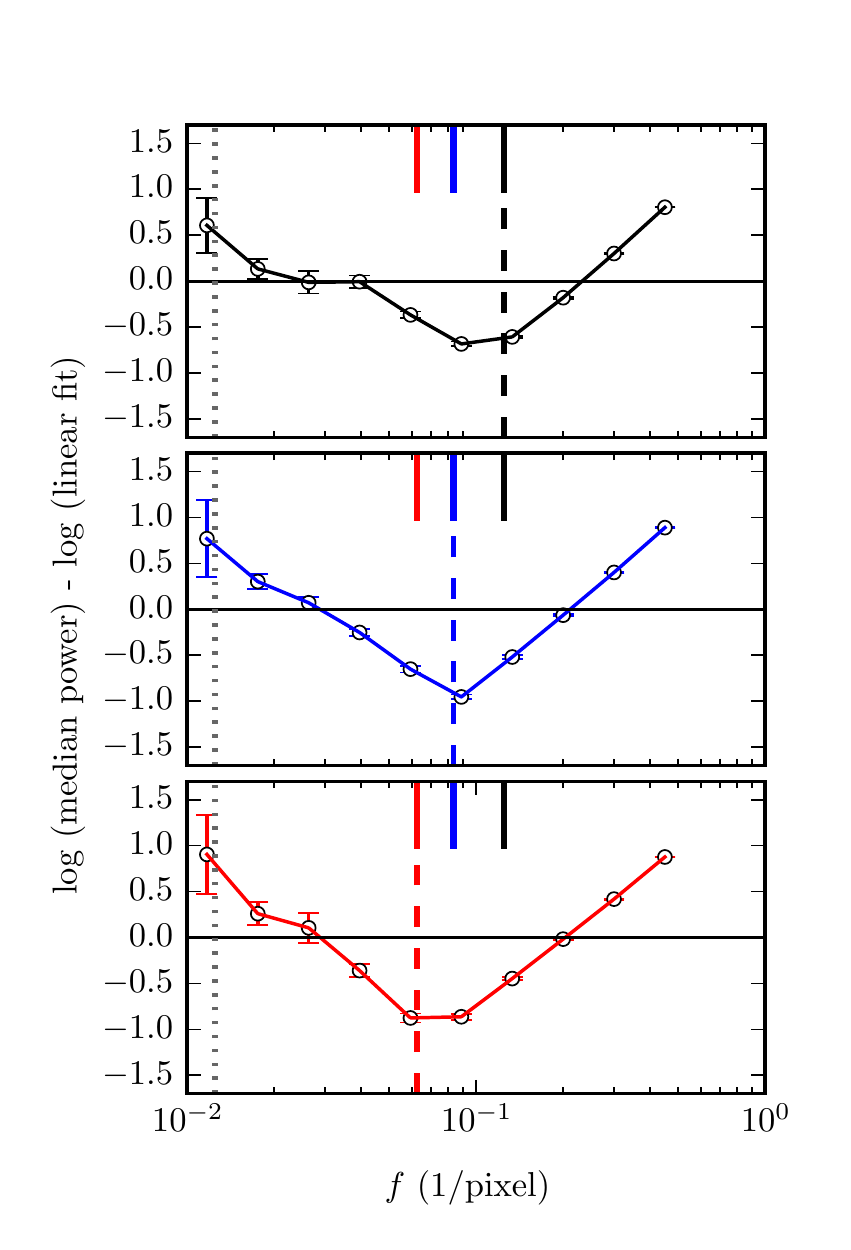}
\caption{Difference between the spatial power spectrum and a linear fit in log-log space (log(median power) - log(fit)) 
for an artificial image with a single filament of constant $\rm R_{flat}$. 
From top to bottom: $\rm R_{flat} = 8, 12, 16$ pixels. The length is 80 pixels for all three images (dotted vertical line).
In each plot, the dashed vertical line shows the spatial frequency that corresponds to $\rm R_{flat}$.
The small vertical lines on the top of each panel show the values of $\rm R_{flat}^{-1}$ of all 3 images, 
for comparison between panels.}
\label{fig:onefilps}
\end{figure}

Fig. \ref{fig:onefilimg} (left) shows a single filament that has $\rm R_{flat} = 8$ pixels and is 80 pixels long. 
The spatial power spectrum of the image (top right) has been fit by a line. The power spectrum deviates significantly
from this line at the spatial frequencies corresponding to $\rm R_{flat}$ (dashed vertical line) and the filament's length 
(dotted vertical line) as can be seen in the bottom left panel of the figure. 

In Fig. \ref{fig:onefilps}, we investigate the effect of changing the width ($\rm R_{flat}$) of this single filament. 
As $\rm R_{flat}$ changes from 8 (top) to 12 (middle) and finally to 16 (bottom) pixels, the signature of this scale
on the power spectrum moves to the corresponding spatial frequencies (indicated by vertical dashed lines).

\begin{figure*}
\centering
\includegraphics[scale=1]{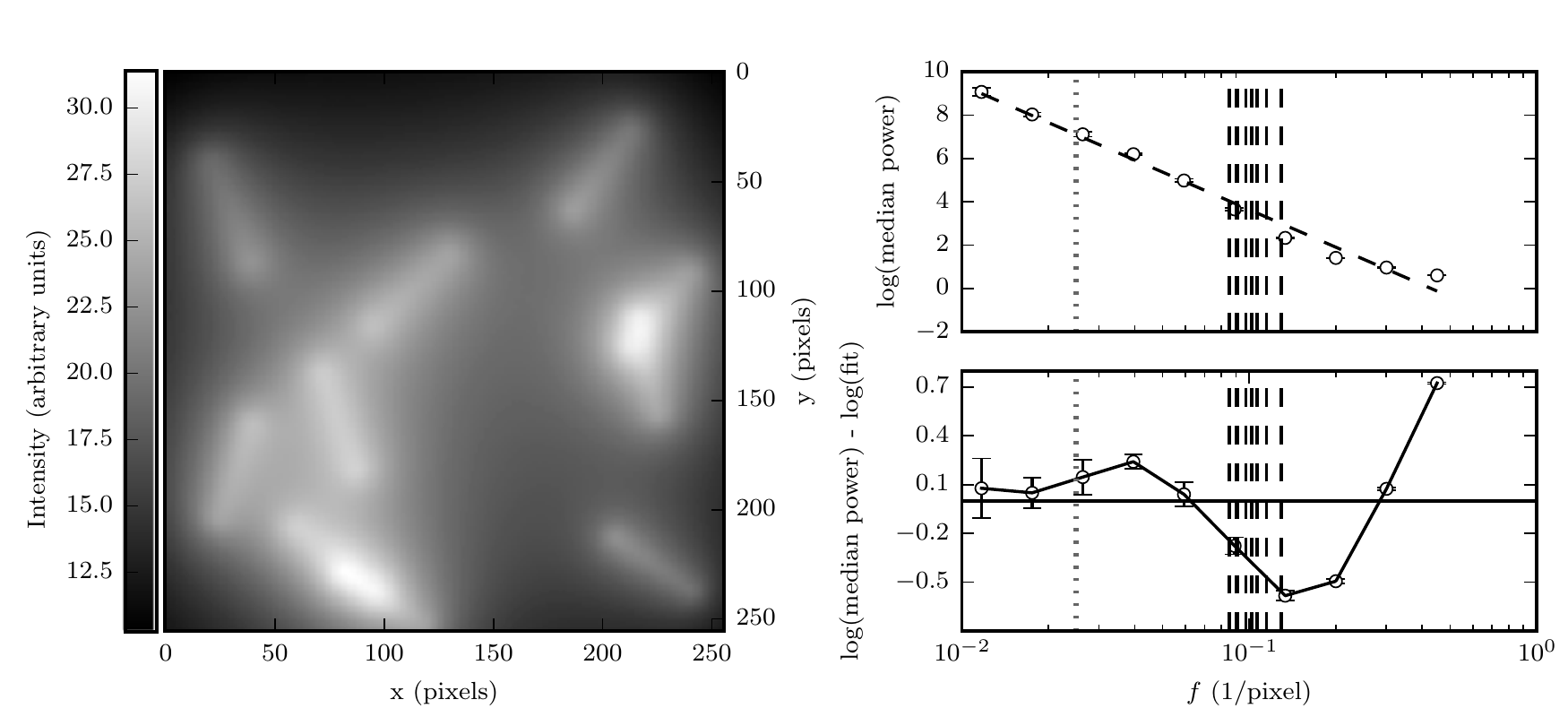}
\caption{Same as Fig. \ref{fig:onefilimg} but for 10 filaments at random orientations, with a length of 40 pixels and 
different $\rm R_{flat}$, drawn from the distribution of mean filament widths in Fig. \ref{fig:all_distr} 
(with values multiplied by 100 so that 0.1 pc is mapped to 10 pixels). The spatial frequency corresponding to the $\rm R_{flat}$ of each of the filaments is shown with a dashed vertical line.}
\label{fig:randfils}
\end{figure*}

Having identified the signature of a characteristic scale on the power spectrum, we now investigate the circumstances under which 
it may be possible to `hide' such a signature in a way that it does not appear in the power spectrum. 
We first create a map with 10 filaments
of the same length (40 pixels) but with random orientations and positions. 
Each filament has a different peak intensity and $\rm R_{flat}$ (constant along its spine). 
The $\rm R_{flat}$ are drawn from the narrow distribution of mean filament widths found in Fig. \ref{fig:all_distr} (bottom panel). 
The values drawn from this distribution were multiplied by 100 to obtain 
$\rm R_{flat}$, meaning that a value of 0.1 pc is mapped to 10 pixels (a scale that is well-sampled in the power spectra of the artificial 
images). The image is shown in Fig. \ref{fig:randfils}, with each of the vertical dashed lines denoting the spatial frequency that 
corresponds to $\rm 1/R_{flat}$ for the 10 filaments. Even when multiple filaments are present, with random orientations and spacings 
between them, the signatures of their characteristic scales are clearly visible as deviations from the fit to the power spectrum. 
These deviations are significantly larger than those seen in the power spectrum of the pink noise. 

We investigate the statistical significance of this result by creating 150 realizations of such images.
We find that in only $\sim10$\% of the images, the maximum residual of the fit is less than 0.2 in amplitude (in the range of frequencies 
corresponding to the values of $\rm R_{flat}$). However, upon visual inspection, these images can be divided into
three categories: (i) the residuals exhibit a systematic offset from the fit, but at a level less than 0.2, 
(ii) only a single point in the power spectrum samples the range of scales used, 
or (iii) most filaments overlap at a certain part of the image, so their individual filamentary structure is not visible.
In the first case, the systematic offset of neighbouring data points is distinct from the random fluctuation of the residuals in the pink 
noise image. Thus, a signature of the width is still identifiable in the spatial power spectrum. 
In the second case, an offset is observed at data points outside but neighbouring to the range considered.
In the last case, the imprint of the larger `cluster' dominates the power spectrum and these images cannot be considered as being 
comprised of filaments.

\begin{figure*}
\centering
\includegraphics[scale=1]{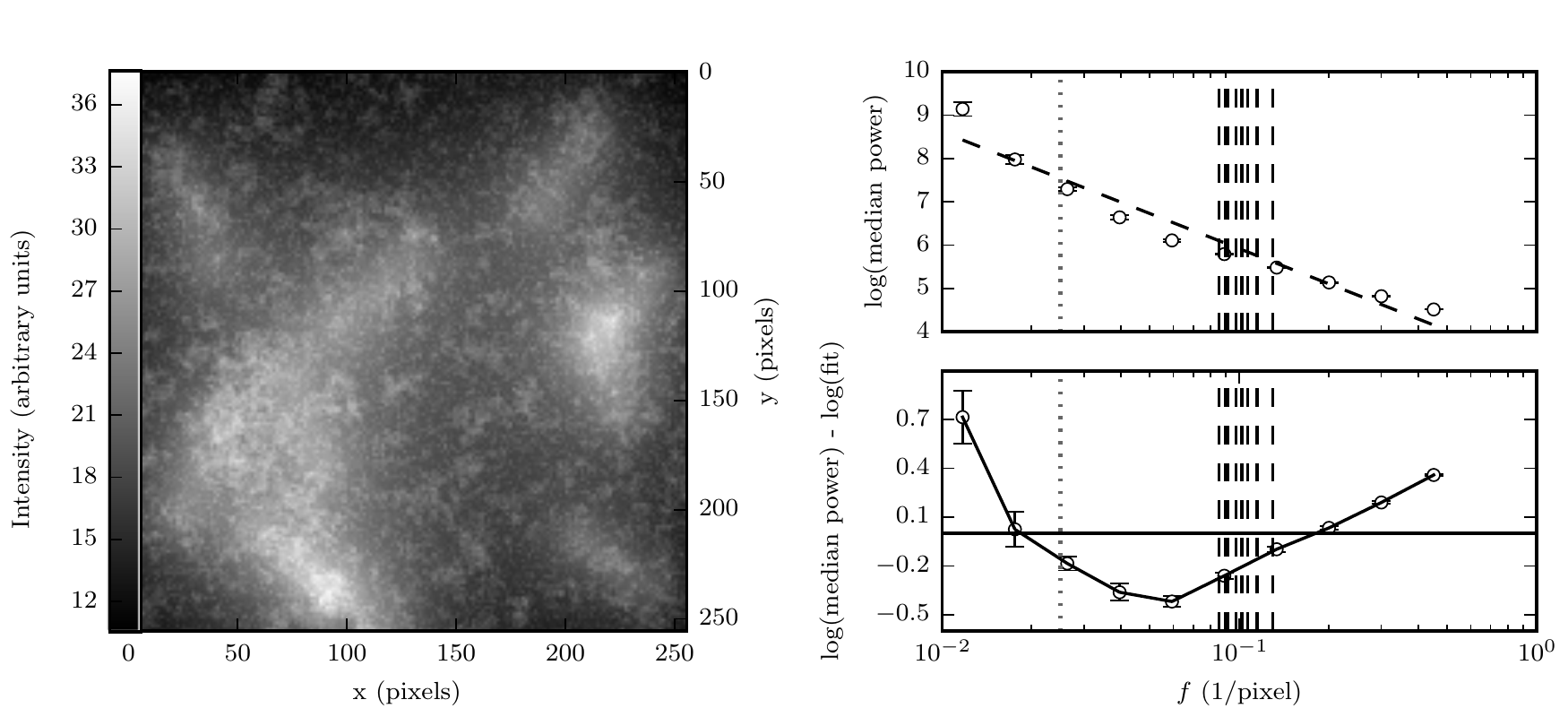}
\caption{Same as Fig. \ref{fig:randfils} but with pink noise with maximum intensity $\sim30$\% that of the highest filament peak.}
\label{fig:noisefils}
\end{figure*}

Finally, we introduce pink noise with very large amplitude (maximum intensity $\sim$30\% that of the largest filament spine 
intensity) in Fig. \ref{fig:noisefils}. The filaments in this image are the same as in Fig. \ref{fig:randfils}.
Now, the signature of the width (a change of the spectral slope at the corresponding spatial frequency range) is not visible 
in the residuals of the power spectrum from the fit. However, the filaments are barely distinguishable from the background noise, which is 
in stark contrast to observations of clouds.

We conclude that the existence of a characteristic scale should appear in the spatial power spectrum of an image 
(e.g. as a change in the spectral slope), provided that the structure is easily discernible from the background.
In the case of the spatial power spectrum of the Polaris Flare \citep{miville}, no such change exists in the power spectrum
at or near the spatial scale corresponding to the ``characteristic'' width (of the very prominent filaments) at 0.1 pc.

\section{ARMA modelling of filament widths} 
\label{sec:appendix}

In section \ref{sec:distros} we saw that the distributions of all profile widths and of filament lengths are
necessary but not sufficient pieces of information to explain the spread of the distribution of mean filament widths 
($\sigma_{mean}$). The goal of this section is to model the variation of widths along the ridge of a filament. 
Using this information, we will be able to explain quantitatively how one obtains a distribution with $\sigma_{mean}$
when averaging the widths from the parent distribution of all profile widths.

We choose to model the variation of profile widths along a filament using an Auto-Regressive-Moving-Average (ARMA) process 
\citep[for a complete description of ARMA modelling, we refer the reader to][]{brockwell}.
Each filament is regarded as a series of $N$ widths, $w_i$, measured at all positions (pixels), $i = 1-N$, along the filament crest.
The difference between the width at point $i$ along the filament crest and the mean width can 
be written as the regression:
\begin{equation}
\delta w_i = \alpha_0 \delta w_{i-1} + \alpha_1 \delta w_{i-2} + \alpha_2 \delta w_{i-3} + ... + \epsilon_i + c_1 \epsilon_{i-1} + ...
\end{equation}
where $\delta w_{i-1}$ is the difference of the width of the point previous to $i$ from the mean width, $\delta w_{i-2}$ is 
measured two points away, and so on. Depending on the order $p$ of the auto-regressive (AR) part of the equation (coefficients $\alpha$),
$\delta w_i$ can have a dependence on the width measured $p$ points away from position $i$.
$\epsilon_i$ is the residual, what is not taken into account by the AR terms. It is assumed random and normally-distributed. 
The terms containing the residuals at different positions 
are the moving-average terms (MA) and their number $q$ is the order of the MA part of the model.

We wish to model the variation of widths along a filament, based on the data in the Polaris Flare.
In order to find an appropriate model, we must first decide on the order of the model to be fit.
The order of the ARMA model ($p$, $q$) can be determined by examination of the Autocorrelation 
and Partial Autocorrelation functions (ACF, PACF) of the widths of filaments in the Polaris Flare 
for $p$ and $q$, respectively \citep[][]{brockwell}. 

The ACF is defined as \citep[following][]{brockwell}:
\begin{equation}
{\rm ACF}(l) = \frac{1}{N \sigma_{fil}^2} \sum_{i=1}^{N-l} (w_{i+l} - \left\langle w \right\rangle) ( w_i - \left\langle w \right\rangle), 
\,0 < l < N
\end{equation}
where $N$ is the number of profiles in a filament, $l$ is the distance (lag) measured along the filament ridge, $w_i$ is the width 
of the $i^{\rm th}$ profile along the ridge and $\left\langle w \right\rangle$ is the average width of the filament. 
Finally, $\sigma_{fil}$ is the standard deviation of profile widths in the filament.
The PACF at a given lag is the autocorrelation at this lag after removal of an AR model of order lag minus 1.
This means that the PACF will be zero at this lag if the AR model effectively removes all correlation.

\begin{figure}
\centering
\includegraphics[scale=1]{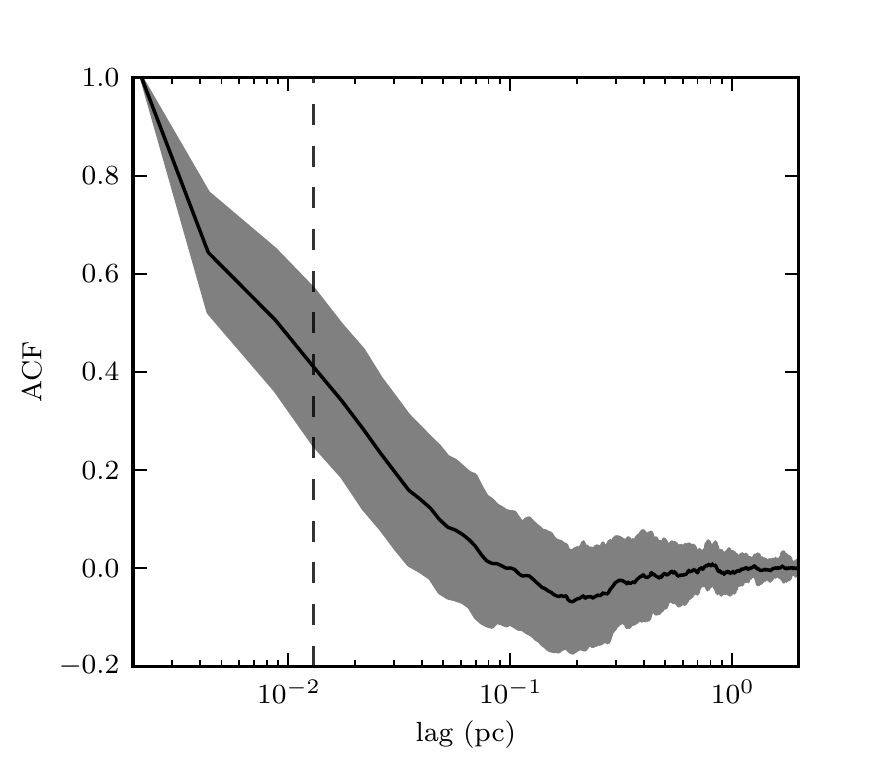}
\caption{ACF of all filaments in the Polaris Flare versus distance. The black line shows the ACF averaged over all filaments for a given 
lag (distance). The $\pm 1$ standard deviation of the ACF of all filaments at a given lag is shown with a grey band.
The vertical line shows the HPBW (beam size) of 0.013 pc. The plot has been truncated at a distance of 2 pc for clarity.}
\label{fig:acf}
\end{figure}

\begin{figure}
\centering
\includegraphics[scale=1]{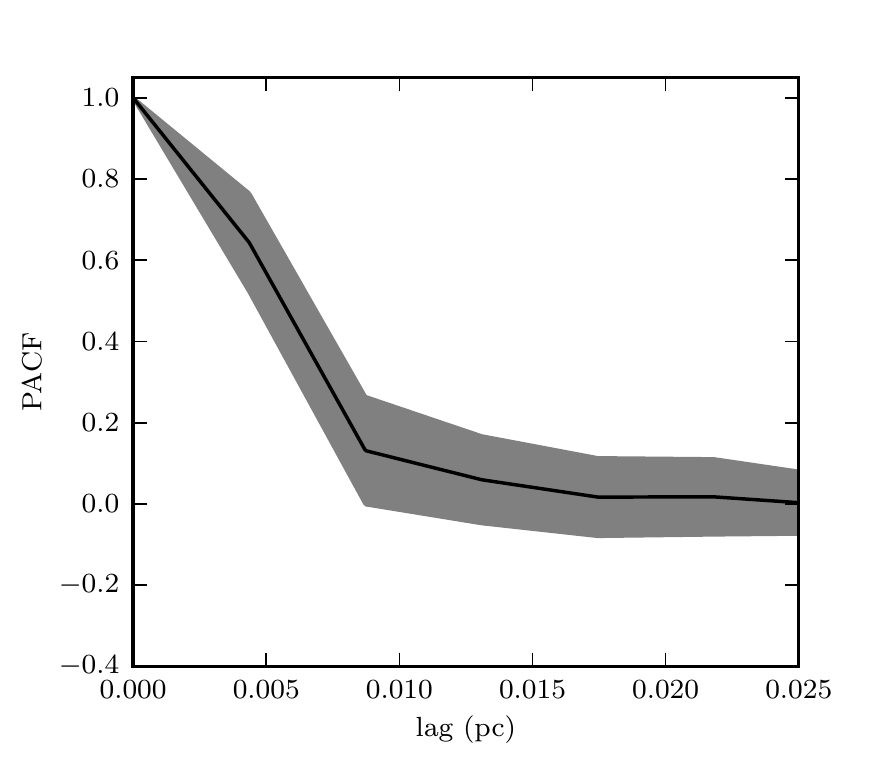}
\caption{As in Fig. \ref{fig:acf} but now showing the PACF of all filaments versus distance. The horizontal axis has been truncated at a 
distance of 0.025 pc.}
\label{fig:pacf}
\end{figure}

Fig. \ref{fig:acf} shows the ACF of widths of all filaments in the Polaris Flare versus distance along the filament ridge\footnote{For 
ARMA modelling and for construction of the ACF we made use of the python module 
{\tt statsmodels} ({\tt http://statsmodels.sourceforge.net/})}.
The ACF drops abruptly and stays around zero for distances larger than approximately twice the beam size 
(HPBW = 0.013 pc, dashed vertical line). For most filaments, widths are strongly correlated only within the beam size.
This corresponds to a lag of 3 pixels, and therefore this is the order of the AR process ($p = 3$). 
The ACF of filaments in IC 5146 and Aquila also follow this trend (for IC 5146 HPBW = 0.04 pc, and for Aquila 
HPBW = 0.023 pc).  
In Fig. \ref{fig:pacf} we plot the PACF of all filaments in the Polaris Flare versus the lag in pc. 
The PACF (averaged over all filaments for every given lag) drops after a lag of 0.004 pc (1 pixel).


Therefore, the (mean-subtracted) widths along filaments in the Polaris Flare can be 
modelled with $p$ = 3 AR terms and $q$ = 1 MA term:
\begin{equation}
\delta w_i = \alpha_0 \delta w_{i-1} + \alpha_1 \delta w_{i-2} + \alpha_2 \delta w_{i-3} + \epsilon_i + c_1 \epsilon_{i-1}.
\label{eqn:arma}
\end{equation}
We fit equation \ref{eqn:arma} to the series of widths of each filament in the Polaris Flare.
We wish to obtain a single model that on average reproduces well the variation of widths along any filament in the cloud.
This model is equation \ref{eqn:arma} where the value of each coefficient is equal to the median of values returned by fitting 
the equation to each filament. The median values of the coefficients from the fits to individual filaments are
$\alpha_1 = 0.8, \, \alpha_2 = 0.2,\, \alpha_3 = 0.05, \, c_1 = -0.23$.

As seen in section \ref{sec:distros}, the observed $\sigma_{mean}$ cannot be explained by only taking into account the distributions of
all profile widths and filament lengths.
We will now estimate what part of $\sigma_{mean}$ can be attributed to the combined contribution of the correlation between the widths of 
neighbouring profiles, and the distribution of filament lengths. We note that the information on the distribution of all profile widths
is included in the ARMA model, as the coefficients were found by fitting to real data.

To this end, we create 100 groups (filaments) of (mean-subtracted) widths in the following way. 
A number of profiles for each filament is drawn from the distribution of filament lengths in the Polaris Flare. 
Each filament is assigned a starting value drawn from a normal distribution with mean 0 and standard deviation
equal to that of the distribution of $\epsilon_i$ from the fits to individual filaments. 
This starting value corresponds to the mean-subtracted width of the first filament profile. 
Consecutive (mean-subtracted) profile widths are found iteratively using equation \ref{eqn:arma}. 
We then calculate the average width of each filament. The distribution of average filament widths has a spread of 0.011 pc, 
similar to the observed $\sigma_{mean}$ of 0.014 pc in the cloud. This process is repeated 100 times 
to quantify whether the difference (0.014 pc versus 0.011 pc) is significant. We find that the 
observed $\sigma_{mean}$ is within the spread of the results of the simulation. 

We have found that $\sigma_{mean}$ can be predicted based on three pieces of information: 
the distribution of all profile widths, the distribution of filament lengths, and the correlation of widths within a beam size.
From the CLT we understand the effect of these three as follows:
A broader distribution of all profile widths will increase the uncertainty on the mean ($\sigma_{mean}$), as seen in equation \ref{eqn:clt}.
From the same equation, it follows that a population of filaments with on average larger lengths will have a smaller $\sigma_{mean}$.
Finally, we have seen that neglecting the effect of the beam (random draws of the width from the parent distribution) produces a 
narrow $\sigma_{mean}$. The effect of the beam is to introduce a larger uncertainty on the mean, by effectively reducing the number
of independent measurements in a single filament.

%%%%%%%%%%%%%%%%%%%%%%%%%%%%%%%%%%%%%%%%%%%%%%%%%%

% Don't change these lines
\bsp	% typesetting comment
\label{lastpage}
\end{document}